# Investigating the theoretical performance of $Cs_2TiBr_6$-based perovskite solar cell with La-doped $BaSnO_3$ and $CuSbS_2$ as the charge transport layers

*Kumar Shivesh[1], Intekhab Alam[2*], A.K. Kushwaha[3], Manish Kumar[4], S.V. Singh[1]*

[1]Department of Chemical Engineering & Technology, Indian Institute of Technology (Banaras Hindu University), Varanasi, Uttar Pradesh-221005, India

[2]Department of Mechanical Engineering, Bangladesh University of Engineering and Technology (BUET), East Campus, Dhaka-1000, Bangladesh

[3]Department of Physics, K.N. Govt. P.G. College, Gyanpur, Bhadohi, Uttar Pradesh-221304, India

[4]Experimental Research Laboratory, Department of Physics, ARSD College, University of Delhi, New Delhi-110021, India

*Corresponding Author
Email: intekhabsanglap@gmail.com, Phone No: +8801819279881

## Abstract

A lead-free, completely inorganic, and non-toxic $Cs_2TiBr_6$-based double perovskite solar cell (PSC) was simulated via SCAPS 1-D. La-doped $BaSnO_3$ (LBSO) was applied as the electron transport layer (ETL) unprecedentedly in the simulation study of PSCs, while $CuSbS_2$ was utilized as the hole transport layer (HTL). wxAMPS was used to validate the results of SCAPS simulations. Moreover, the first-principle density function theory (DFT) calculations were performed for validating the 1.6 eV bandgap of the $Cs_2TiBr_6$ absorber. To enhance the device performance, we analyzed and optimized various parameters of the PSC using SCAPS. The optimum thickness, defect density, and bandgap of the absorber were 1000 nm, $10^{13}$ cm$^{-3}$, and 1.4 eV, respectively. Furthermore, the optimum thickness, hole mobility, and electron affinity of the HTL were 400 nm, $10^2$ cm$^2$V$^{-1}$s$^{-1}$, and 4.1 eV, respectively. However, the ETL thickness had a negligible effect on the device's efficiency. The optimized values of doping density for the absorber layer, HTL, and ETL were $10^{15}$, $10^{20}$, and $10^{21}$ cm$^{-3}$, respectively. Herein, the effect of different HTLs was analyzed by matching up the built-in voltage ($V_{bi}$) in respect of the open-circuit voltage ($V_{OC}$). It was found that the $V_{bi}$ was directly proportional to the $V_{OC}$, and $CuSbS_2$ was the champion in terms of efficiency for the PSC. The optimum work function of metal contact and temperature of the PSC were 5.9 eV and 300 K, respectively. After the final optimization, the device achieved an exhilarating PCE of 29.13%.

# 1. Introduction

The halide perovskites (HPs) possess the structure of $ABX_3$, where A represents a monovalent cation, B stands for a divalent metal cation, and X denotes a halide anion [1]. These are a group of semiconductor materials with atypical optoelectronic properties, like high absorption coefficient, longer carrier diffusion length, weakly bound exciton, and a wide range of bandgap tunability [2], [3]. However, although these light-harvesting layers surpassed the power conversion efficiency (PCE) in perovskite solar cells (PSCs), there are two leading problems: the toxicity of lead and the PSCs that give high PCE consist of organic cations, like methylammonium and formamidinium ions [4]. These ions are extremely volatile and hygroscopic, thus making the device unstable and more intolerant of heat and moisture [5].

Ti-based $A_2^{+1} Ti^{+4} X_6^{-1}$ double perovskites are lead-free and non-toxic materials that not only can substitute the lead in perovskite but also possess novel optoelectronic applications besides solar cells [6]. According to the density functional theory (DFT) calculations, these HPs have advantageous electronic and optical properties, such as appropriate bandgaps and tuneable defect properties. The processability and stability of these HPs are encouraging their utilization in PSCs [7]. In 2017, $Cs_2TiX_6$ (X = F, Cl, Br, or I) was found as an excellent option for an eco-friendly, earth-abundant, and reasonable perovskite light-absorbing layer. The tunability of the bandgap from 1.4 eV to 1.8 eV makes these perovskites fit for the implementation in solar cells [8]. In addition, these HPs have satisfactory electron ($L_n$) and hole ($L_p$) diffusion lengths [9]. In the present work, $Cs_2TiBr_6$ was selected as the absorber layer of the PSC. It exhibits prolonged carrier-diffusion lengths, efficient photoluminescence, as well as energy levels appropriate for the usage in tandem solar cells [6]. Besides, it demonstrates superior thermal, moisture, and light effects stability than the methylammonium lead iodide (MAPbI$_3$) thin films, ensuring the higher intrinsic or environmental durability of the Cs$_2$TiBr$_6$ thin films [7]. Moreover, Cs$_2$TiBr$_6$-based PSC has attained 3.3% efficiency experimentally that is greater than most other double PSCs [7].

The valence band offset (VBO) between the hole transport layer (HTL) and absorber, hole mobility, and cost are the major properties for consideration while selecting an HTL for the PSCs [10]. The researchers mostly use the organic spiro-OMeTAD as the HTL in the PSCs because of its good tunability and processability [2]. However, its poor conductivity, very low hole mobility, high fabrication cost, and environmental instability obstruct its application [11]. Therefore, the inorganic p-type semiconductors are better replacements as the HTL because they possess outstanding chemical stability along with excellent hole mobility [12]. Especially, inorganic Cu-based chalcogenide compounds denoted as Cu$_a$BX$_b$, where X = Te, S, or Se and B = Bi, Sn, or Sb, are under investigation for their application in the photovoltaics. The most abundant and cost-effective member of this family is CuSbS$_2$ [13], [14]. It possesses a direct bandgap of 1.58 eV along with an excellent hole mobility of 49 cm$^2$V$^{-1}$s$^{-1}$ [15]. Moreover, its band alignment with Cs$_2$TiBr$_6$ perovskite is pretty good for the charge carriers' transfer, thus making it an appropriate HTL for this work.

Although most PSCs use TiO$_2$ as the electron transport layer (ETL), their stability declines under ultraviolet illumination due to the presence of mesoporous-TiO$_2$ [16]. To make PSCs more stable, different approaches have been tried, such as fabricating ETL-free PSCs, introducing interfacial layer between perovskite layer and ETL, doping of TiO$_2$, use of filter UV photocatalytic ability, and insertion of a down-converting layer [17], [18]. Nevertheless, there is another way, which is the replacement of mesoporous-TiO$_2$ with a new ETL without deteriorating the PCE of the PSC. BaSnO$_3$ (BSO), a broad bandgap n-type perovskite oxide, exhibits extensive applications, such as a gas sensor [19], thin-film transistor [20], and transparent conducting oxide [21]. Specifically, La-doped BaSnO$_3$ expressed as LBSO has high electrical mobility at the ambient temperature [22]. It exhibits lesser ultraviolet photocatalytic ability because of its minimal dipole moment, attributed to the cubic structure without the most frequently occurring distortion in perovskites, i.e., octahedral tilting [23]. In addition, the conduction band offset (CBO) between the LBSO and Cs$_2$TiBr$_6$ perovskite is quite low, which makes it a suitable match for this work. The device architecture of the first experimentally fabricated PSC with LBSO as the ETL was FTO/LBSO/MAPbI$_3$/PTAA/Au that had a PCE of 21.3% [23].

In the present work, we reported LBSO as an ETL in the simulation investigations for the first time and designed a fully inorganic and non-toxic FTO/LBSO/Cs$_2$TiBr$_6$/CuSbS$_2$/Au heterostructure via SCAPS-1D simulator. We analyzed and optimized the effect of change in thickness, defect density, doping density, and bandgap of the absorber layer, thickness and doping density of ETL along with thickness, doping density, electron affinity, and hole mobility of HTL. We also optimized the operating temperature and metal work function of the PSC. Furthermore, we validated the simulation results using the wxAMPS software. In addition, we also performed theoretical first-principle DFT calculations to validate the bandgap of the Cs$_2$TiBr$_6$ perovskite absorber.

## 2. Methodology

### 2.1 SCAPS and wxAMPS simulation methodology

The 1-D simulation software SCAPS was used to carry out the numerical study for this work. This C language-based program has been developed by Professor M. Burgelman at the University of Gent, Belgium [24]. This software uses some fundamental semiconductor equations, which are shown in the Supplementary File, to calculate the external quantum efficiency (EQE) curves, current density-voltage (J-V) curves, energy bands, and ac characteristics [3].

We validated our SCAPS simulation results using a 1-D solar simulator software wxAMPS. wxAMPS has been developed at the UIUC, Illinois, United States [2]. Actually, it is an updated version of AMPS developed at the PSU, Pennsylvania, United States [25]. It can generate EQE curves, J-V curves, electrical field distribution, as well as carrier concentration and recombination profiles utilizing some fundamental semiconductor equations [26]. The relevant equations, as well as the comparative results, have been depicted in the Supplementary File (Table S1 and Fig. S1) [4]. The simulation results of wxAMPS dictated that the values of the photovoltaic properties, namely open-circuit voltage ($V_{OC}$), fill factor (FF), short-circuit current ($J_{SC}$), and PCE, were very similar to the values obtained from SCAPS simulation for the PSC. In addition, the SCAPS simulation working procedures are also represented in the Supplementary File (Fig. S2).

**2.2 Device architecture**

In this present work, we proposed a device structure of FTO/LBSO/$Cs_2TiBr_6$/$CuSbS_2$/Au, as shown in Fig. 1(a). The band energy alignment diagram and band energy diagram of the device have been demonstrated in Fig. 1(b) and Fig. 1(c), respectively. $CuSbS_2$ was utilized as the p-type inorganic HTL, and $Cs_2TiBr_6$ was used as the perovskite absorber layer. For the very first time, LBSO was used as the ETL for the numerical simulation study of PSC structure. Fluorine-doped tin oxide (FTO) and gold (Au) were used as front contact and back contact, respectively.

**2.3 Simulated parameters**

The data of electrical and optical properties for different layer materials were collected from the previously published experimental and computational research articles. The basic parameters for charge transport layers and absorber layer are tabulated in Table 1. The bulk defect properties of all the layers are listed in Table 2. To make the device closer to reality, we introduced two interfacial defect layers (IDLs) in the PSC. The first one was placed between ETL/perovskite, and another one was inserted between perovskite/HTL [27]. Table 3 shows the interface defect properties of the inserted IDLs. The metal work function ($\phi_{BC}$) of Au was taken as 5.1 eV, and $10^7$ cms$^{-1}$ was assumed as the thermionic emission/surface recombination velocities for both electrons and holes. All the calculations had been executed at 300 K along with the standard AM1.5G spectrum [28]. The "Eg-sqrt" model had been employed to set the optical absorption constant, α(hv), for each layer. The description of this model has been exhibited in the Supplementary File [2].

**2.4 First-principle DFT study of $Cs_2TiBr_6$**

For getting a better insight into the electronic behavior of the studied $Cs_2TiBr_6$ compound, its electronic band structure and partial density of states (DOS) had been calculated utilizing DFT. The unit cell of the $Cs_2TiBr_6$ structure was generated via VESTA (a 3D visualization software) and has been shown in Fig. 2(a). All the DFT calculations were done by QUANTUM ESPRESSO distribution [29]. The calculations were done under ultrasoft pseudopotential within the framework of GGA-PBE exchange correlation functional [30], [31]. The kinetic energy cut-offs for wave functions were used 600 Ry. For the Brillouin zone sampling of electronic states, we used a k-point mesh of 10 x 10 x 10 Monkhorst-Pack [32]. The calculated band structure is illustrated in Fig. 2(b), showing that there was a bandgap of 1.61 eV between the valence band maxima (VBM) and the conduction band minima (CBM). This result was very much comparable with the previously calculated results and the 1.6 eV bandgap of our absorber [9], [33]. We also calculated the partial DOS for the titled compound, and the results are depicted in Fig. 2(c). Fig. 2(c) represents that the valence band between the energy range of -6 eV to -3 eV was mainly constructed by the hybridization of the electrons of 6s-orbitals of Cs-atoms and the 3d-orbitals of Ti-atoms, while the formation of valence band near the Fermi surface was mainly due to the hybridization of 3d-orbitals of Ti-atoms and the 5p-orbitals of Cs-atoms. Besides, the minima of the conduction band had been mainly constructed by the hybridization of 3d-orbitals of Ti-atoms and the 6s-orbitals of Cs-atoms. The higher energy side of the conduction band was mainly constructed by the hybridization of 3d-orbitals of Ti-atoms and the 4p-orbitals of Br-atoms.

## 3. Results and Discussion

### 3.1 Effect of the change in absorber layer thickness

In the PSCs, the light-absorbing layer plays a pivotal role in deciding the PCE. In order to find the optimum thickness of the perovskite layer, the thickness of $Cs_2TiBr_6$ was varied from 700 nm to 2500 nm without changing other parameters. The change in $V_{OC}$, $J_{SC}$, FF, and PCE was observed while varying the absorber thickness, and the trends have been illustrated in Fig. 3(a-d).

The $V_{OC}$ increased from 1.0262 V to 1.0274 V with the increasing absorber thickness because the photon-capturing ability of the absorber layer rises with its thickness, which enhances the rate of generation of the charge carriers [27]. Furthermore, the $J_{SC}$ improved from 24.2399 mA/cm$^2$ to 25.2551 mA/cm$^2$ with the augmentation of absorber thickness. Because there will be a rise of the spectral response at the longer wavelength with the increasing thickness [34]. However, the FF dropped monotonically from 84.70% to 79.03% with the increasing absorber thickness; the deterioration is because of the enhanced series resistance [35]. This can be because of the superiority of carrier recombination along with the presence of parasitic resistance losses [36]. The PCE increased initially with the thickness and reached a maximum value of 21.21% at 1000 nm, which was chosen as the final optimum absorber thickness. Then it declined with a further enhancement in the absorber thickness. The obtained optimum absorber thickness was slightly higher than the 800 nm optimum thickness found for the same absorber material in the published work [37]. The initial rise of PCE can be attributed to an increment in the generation of electron-hole pairs with the increasing thickness. However, the drop of PCE at the higher absorber thickness is due to the enhanced radiative recombination and charge pathway resistance [4].

**3.2 Effect of the change in ETL thickness**

For designing highly efficient PSCs, the parameters of charge transport layers should be carefully chosen. An appropriate ETL helps in decreasing the recombination currents and increasing the transmittance in PSCs [38]. To find the optimized performance of the PSC, the thickness of LBSO ETL was modulated from 50 nm to 500 nm while maintaining other parameters constant. The effect of ETL thickness on the $V_{OC}$, $J_{SC}$, FF, and PCE has been illustrated in Fig. S3(a-d) of the Supplementary File.

The $V_{OC}$, FF, $J_{SC}$, and PCE remained almost invariable throughout the variation. The very marginal decrease in $J_{SC}$ and PCE can be due to the partial absorption of light by a thicker ETL, which results in a reduction of the rate of charge generation and collection [39]. The relation between the LBSO thickness and transmittance can be shown by the following equation [40]:

$$\alpha = \frac{1}{d} \ln \frac{1}{T} \dots \dots \dots \dots (1)$$

Where, α is the absorption coefficient, d is the film thickness, and T is the transmittance. It has been reported that the performance loss due to the increasing thickness of ETL gets higher with the reduction in transmittance [39]. It could be concluded from our result that the thickness of LBSO had a negligible effect on PSC performance similar to the published literature [41]. So, 120 nm was taken as the optimum ETL thickness, which was equal to the experimental work [23].

**3.3 Effect of the change in HTL thickness**

To optimize the thickness of $CuSbS_2$ HTL, the photovoltaic parameters were determined between 100 nm and 1000 nm. The impact of HTL thickness on the $V_{OC}$, $J_{SC}$, FF, and PCE has been illustrated in Fig. S4(a-d) of the Supplementary File.

A similar increasing trend of $V_{OC}$, $J_{SC}$, and PCE was noticed with the increasing $CuSbS_2$ thickness. The $V_{OC}$, $J_{SC}$, and PCE enhanced from 0.9871 V to 1.0506 V, 24.4069 mA/cm$^2$ to 25.1296 mA/cm$^2$, and 20.55% to 21.67%, respectively. But the FF monotonously reduced from 85.30% to 82.07% as the HTL thickness enhanced. In the previous section, it was observed that the PCE decreased very slightly with the increasing ETL thickness. However, the PCE increased with the increasing HTL thickness similar to the published literature [42]. To minimize the chances of recombination, generally, a p-type layer should be thicker compared to an n-type layer. Because it helps in transporting an equal number of charge carriers to the terminal immediately [43]. So, the optimum thickness of HTL should be greater than the optimum thickness of ETL, i.e., 120 nm. Therefore, the optimum thickness of HTL was taken 400 nm as the improvement in PCE after that was very marginal. The increased thickness of HTL reinforces the absorption of photons on the light-harvesting layer [43].

**3.4 Effect of the change in absorber layer defect density**

The absorber defect density ($N_t$) has a vital influence in deciding the efficiency of PSCs. The PCE of solar cells is greatly influenced by the morphology and quality of the light-absorbing layer. The irradiation of light on the perovskite absorber layer generates photoelectrons. However, poor morphology can cause inadequate coverage of the perovskite layer on the ETL. The inferior quality of film results in higher defect density, which in turn, causes higher recombination [12]. The $N_t$ was modulated from $10^9$ cm$^{-3}$ to $10^{16}$ cm$^{-3}$ for this section. It can be noticed from Fig. 4(a-d) that all of the photovoltaic parameters decreased with the increasing defect density. The $V_{OC}$, $J_{SC}$, FF, and PCE significantly reduced from 1.0266 V to 0.8190 V, 24.73 mA/cm$^2$ to 23.0209 mA/cm$^2$, 87.40% to 48.09%, and 22.19% to 9.07%, respectively.

The PSC performance remained almost constant up to $10^{12}$ cm$^{-3}$ absorber defect density. When the defect density value exceeded $10^{12}$ cm$^{-3}$, the performance of the PSC started to decline. This declination can be ascribed to the non-radiative Shockley-Read-Hall (SRH) recombination, which is a principal cause for lifetime reduction, carrier recombination, and a significant reduction in the device performance [2]. The corresponding equations of SRH recombination have been described in the Supplementary File [4]. To minimize the defect densities, a perovskite with high crystallinity can be an option. The crystallinity can be improved with the help of pertinent conditions of layer processing [35]. At the lower defect density, the PSC showed higher PCE. But we couldn't take the $N_t$ of $10^9$ cm$^{-3}$ as the optimum value at which the PCE was maximum. Because it is not possible to synthesize a material with such a small $N_t$ experimentally [44]. Thus, the optimum defect density of $Cs_2TiBr_6$ was taken to be $10^{13}$ cm$^{-3}$ without sacrificing much performance. This optimum absorber $N_t$ was slightly lower than the $10^{14}$ cm$^{-3}$ optimum $N_t$ attained for the same absorber layer in the published literature [45]. Moreover, the modulation in the total recombination profile of the solar cell with various absorber defect densities has been plotted in Fig. S5 of the Supplementary File.

**3.5 Effect of the change in absorber layer doping density**

To understand how the acceptor doping concentration ($N_A$) of the absorber layer affects the photovoltaic parameters, the $N_A$ of the $Cs_2TiBr_6$ layer was varied from $10^9$ cm$^{-3}$ to $10^{19}$ cm$^{-3}$. Fig. 5(a-d) illustrates the modulation of $V_{OC}$, $J_{SC}$, FF, and PCE with different absorber doping densities. The $V_{OC}$ and $J_{SC}$ improved from 1.0258 V to 1.0267 V and 24.6214 mA/cm$^2$ to 24.7282 mA/cm$^2$, respectively, whereas the FF and PCE dropped from 87.70% to 83.53% and 22.15% to 21.21%, respectively. The PV parameters were constant up to $10^{17}$ cm$^{-3}$ $N_A$, which signifies that under the incident of the same number of photons, the generation rate of photo-generated carriers is constant with the absorber doping density.

The Fermi energy level of the hole drops when the absorber doping density increases, which causes the $V_{OC}$ to rise [44]. The built-in potential increases with the increasing $N_A$ of the absorber. This can be another reason behind the increment in $V_{OC}$ because of the elevation of charge separation. Nevertheless, when the $N_A$ value exceeded $10^{17}$ cm$^{-3}$, the PCE decreased. As the defect states start to increase at higher $N_A$, the PCE falls [44]. The optimum $N_A$ lied between $10^9$ cm$^{-3}$ and $10^{17}$ cm$^{-3}$ as the $V_{OC}$, $J_{SC}$, FF, and PCE were constant in this range [46]. The generation of an electric field at the interface layers of the PSC occurs with the increasing absorber doping density. Nevertheless, the occurrence of the recombination of charge carriers is also possible with the generation of the electric field. Therefore, to achieve superior performance, an optimum value of acceptor doping density should be chosen [47]. As the optimum value of $N_A$ should be moderate, it was taken to be $10^{15}$ cm$^{-3}$, which was nearly equal to the published research work [12].

**3.6 Effect of the change in acceptor doping density of HTL**

There are two possible approaches by which the doping of ETL and HTL can be accomplished. It can be achieved with minority carriers, but this approach drastically drops the photovoltaic parameters. Conversely, it can also be accomplished by majority carriers that enormously enhance the PV parameters. An intermediate level of doping density will be helpful for obtaining a better performance of PSCs [34]. To understand the impacts of acceptor doping concentration ($N_A$) of HTL, the $N_A$ of the CuSbS$_2$ layer was augmented from $10^{17}$ cm$^{-3}$ to $10^{21}$ cm$^{-3}$ [41]. Fig. S6(a) shows the PCE with the varying $N_A$ of HTL, while Fig. S6(b) depicts the J-V curves with different $N_A$ values of HTL in the Supplementary File.

The $V_{OC}$ enhanced from 0.9389 V to 1.1768 V with the increasing $N_A$ of HTL. The increment in the built-in electric potential at the HTL/perovskite interface is the reason behind the higher value of $V_{OC}$ at higher $N_A$ [26]. Besides, the $J_{SC}$ and PCE improved from 24.6859 mA/cm$^2$ to 24.7464 mA/cm$^2$ and 19.7% to 21.88%, respectively, with the augmentation of $N_A$ of HTL. The interface electric field among the layers of PSC increases at higher $N_A$, which brings about the enhancement of electric potential. This reinforces the separation of the charge carriers with reduced recombination speed and enhances the PCE of the PSC [48]. The optimum $N_A$ of CuSbS$_2$ was taken to be $10^{20}$ cm$^{-3}$ instead of $10^{21}$ cm$^{-3}$ at which the maximum PCE was attained. Because a higher value of $N_A$ can create deep coulomb traps, thus declining the hole mobility [41]. This optimum $N_A$ of HTL was comparable to the previously published literature [49].

**3.7 Effect of the change in donor doping density of ETL**

Fig. 6(a-d) illustrates the variation of main photovoltaic parameters of the PSC, namely the $V_{OC}$, $J_{SC}$, FF, and PCE, against the doping concentration ($N_D$) of LBSO. To find the optimum doping concentration of ETL, the $N_D$ of LBSO was varied from $1 \times 10^{15}$ cm$^{-3}$ to $1 \times 10^{21}$ cm$^{-3}$.

The $V_{OC}$ was constant throughout the variation, whereas the $J_{SC}$ narrowly decreased from 24.73 mA/cm$^2$ to 24.7282 mA/cm$^2$ by increasing the $N_D$ of ETL. However, the FF and PCE slightly enhanced from 83.22% to 83.53% and 21.13% to 21.21%, respectively, with the increasing $N_D$, respectively. Because a higher value of the $N_D$ of ETL facilitates charge extraction and charge transport at the ETL/perovskite interface. The performance of the PSC deteriorated at the lower doping density of LBSO, which is attributed to the high series resistance [41], [50]. The optimum $N_D$ of the ETL was taken corresponding to the maximum PCE, i.e., $1 \times 10^{21}$ cm$^{-3}$ for the PSC, which was nearly equal to the $2 \times 10^{21}$ cm$^{-3}$ $N_D$ of the experimental work [51].

**3.8 Effect of the change in absorber bandgap**

The energy gap ($E_g$) of the absorber layer has a major impact on the PCE of the PSC. The bandgap tunability is the most unique property of the perovskites. In this section, the perovskite energy gap was modulated from 1.2 eV to 1.6 eV. Fig. 7(a-d) shows the trends of $V_{OC}$, $J_{SC}$, FF, and PCE versus the absorber layer energy gap.

The $V_{OC}$ and FF increased steadily from 0.8549 V to 1.0267 V and 59.80% to 83.53%, respectively, with the increasing absorber bandgap [27]. But there was a drastic decrement in $J_{SC}$ from 38.0699 mA/cm$^2$ to 24.7282 mA/cm$^2$ with an increase in the energy gap. Because at a high energy gap, fewer photons get absorbed due to their lower energy compared to the energy gap [41]. There was an increment in PCE from 1.2 eV to 1.4 eV bandgap, achieving a maximum value of 21.87%. However, the PCE started declining after the 1.4 eV bandgap as there is a trade-off. If the bandgap is too high, then an insufficient number of electrons generate due to the very high energy of few photons. In contrast, too low an energy gap enhances the number of electrons but the majority of the energy is dissipated as heat [52]. So, a higher or a lower absorber bandgap with respect to the ideal value of 1.4 eV makes the device inappropriate for solar cell applications due to its declined sunlight absorption capabilities. So, we took 1.4 eV as the optimum absorber bandgap that also exhibited the maximum PCE, similar to the published literature [53].

**3.9 Effect of the change in electron affinity of HTL**

To enhance the PCE of a PSC, an appropriate energy level alignment between the HTL and perovskite layer is a decisive step. For this reason, the valence band offset (VBO = $E_{V, Absorber}$ - $E_{V, HTL}$) between the HTL and the perovskite layer needs to be analyzed, where $E_V$ is the valence band energy level. Moreover, for balancing among the photovoltaic parameters, the VBO should be less than zero [54]. In this study, the electron affinity ($\varkappa_e$) of CuSbS$_2$ was varied from 3.8 eV to 4.3 eV with respect to the perovskite layer for investigating the effect of VBO. The effect of $\varkappa_e$ and $E_V$ of CuSbS$_2$ on the PSC performance has been shown in Fig. S7 of the Supplementary File.

The observed trends in $V_{OC}$ and PCE with varying $\varkappa_e$ of HTL were almost identical to the published literature [54]. When the VBO was increased from -0.69 eV to -0.39 eV, the $V_{OC}$ and PCE significantly improved from 0.879 V to 1.119 V and 17.11% to 22.4%, respectively. After

-0.39 eV VBO, we observed a drastic reduction in $V_{OC}$ and PCE. Therefore, the optimum VBO for the PSC was taken to be -0.39 eV (corresponding to 4.1 eV $\varkappa_e$ and -5.68 eV $E_V$ of $CuSbS_2$) that gave the maximum PCE. To attenuate the charge recombination at the HTL/perovskite interface, the reduction of VBO is desirable. The deep lowest unoccupied molecule level of HTL with respect to the perovskite layer enables the efficiency improvement of solar cells [54]. After the optimum VBO of -0.39 eV, the PCE dropped that can be because of the creation of the Schottky barrier for the holes. Because of its formation, the propagation of holes to the back contact will be hindered, causing the reduction of the PCE of the PSC [5].

**3.10 Optimization of the HTL**

In this work, we compared the performance of four different HTLs, namely CuO, $Cu_2O$, PEDOT:PSS, and P3HT, with the $CuSbS_2$ HTL. It assisted us to find out the most suitable HTL for our proposed PSC. We kept the properties of the ETL and perovskite layer unchanged and simulated for different HTLs. For the sake of proper comparison, we also fixed all the bulk and interface defect panel properties the same as the initial device. Moreover, the initial values of thickness, effective density of states, and $N_A$ of $CuSbS_2$ HTL were used for all the HTLs. For the appropriate propagation of electrons from the perovskite to ETL, the CBM of ETL should position below the CBM of the absorber layer. Likewise, for the smooth transfer of holes from the perovskite to HTL, the VBM of HTL should position above that of the perovskite layer. The HTLs were optimized in this study by correlating the $V_{OC}$ to the built-in potential ($V_{bi}$), where $V_{bi}$ is the difference between the conduction band energy level between the perovskite and ETL interface to the perovskite and HTL interface ($E_{C\_PVK/ETL} - E_{C\_PVK/HTL}$) divided by the elementary charge (q) [55]. The basic parameters for the alternative HTLs are tabulated in Table 4. Besides, the simulated parameters for various HTLs have been tabulated in Table 5.

Fig. 8(a-b) shows the photovoltaic parameters, whereas Fig. 9(a) and Fig. 9(b) illustrate J-V and QE curves, respectively. Fig. 1(b), as well as Fig. S8(a-d) of the Supplementary File, display the energy band alignment diagrams. Furthermore, Fig. 1(c), as well as Fig. S9(a-d) of the Supplementary File, display the energy band diagrams.

From Table 5, it was apparent that the $V_{bi}$ was directly proportional to the $V_{OC}$ similar to the published work [55]. The lowest PCE of 12.82% was shown by P3HT HTL because of the $E_{C\_ETL}$-$E_{V\_HTL}$ = 0.55 eV, which exhibited the least $qV_{bi}$ of 0.29 eV and thus dropped the $V_{OC}$ to 0.729 V. Moreover, P3HT had the lowest charge carrier mobility among the investigated HTLs, and its conductivity showed non-linear characteristics with respect to the doping concentration [55]. Because of these reasons, the P3HT HTL had the worst performance among all the HTLs. However, the $CuSbS_2$ HTL gained the highest $qV_{bi}$, i.e., 0.91 eV, which increased the $V_{OC}$ to 1.0267 V. It was the primary reason for the highest PCE of 21.21% achieved by the $CuSbS_2$ HTL. Furthermore, $CuSbS_2$ had a good charge carrier mobility and superior band alignment with the perovskite layer and metal contact. Hence, the primarily used $CuSbS_2$ was considered as the best HTL for the PSC. Moreover, there was not much variation in the QE curves for different HTLs. Because the optical absorption coefficient of the HTL is insignificant as it is positioned at the rear end of the device [56].

**3.11 Effect of the change in back contact work function**

The creation of an ohmic contact is mandatory for facilitating the appropriate collection of holes through the back contact [55]. In our previous simulations, Au was employed as the back contact with a work function ($\phi_{BC}$) of 5.1 eV. Different $\phi_{BC}$ values were analyzed in this section to understand their effect on the PSC. Fig. 10(a-d) illustrates the effect of $\phi_{BC}$ on the $V_{OC}$, $J_{SC}$, FF, and PCE, while the $\phi_{BC}$ was tuned from 4.9 eV to 6.0 eV. Fig. 11(a) illustrates that when the $\phi_{BC}$ was 4.9 eV, a Schottky barrier was present in the energy band diagram, whereas there was no barrier when the $\phi_{BC}$ was 6.0 eV according to Fig. 11(b). A Schottky barrier for holes can be generated when the work function becomes equal to the VBM of HTL or lower than the VBM of HTL as depicted in Fig. 11(a). But the Schottky barrier can vanish at a higher value of $\phi_{BC}$ as the $\phi_{BC}$ matches with the Fermi level of HTL as shown in Fig. 11(b), which results in the improvement of PCE.

The observed trends in $V_{OC}$, $J_{SC}$, FF, and PCE with varying $\phi_{BC}$ were similar to the published literature [57]. The $V_{OC}$ and $J_{SC}$ enhanced from 0.8267 to 1.2194 V and 24.727 to 25.0959 mA/cm$^2$, respectively, with the increasing $\phi_{BC}$. When the $\phi_{BC}$ is low, the $V_{OC}$ falls because of the declination in the built-in voltage of the PSC. At the same time, the $J_{SC}$ decreases because of the inadequate collection of electron-hole pairs [2]. Moreover, with the increasing $\phi_{BC}$, the FF increased to a maximum value of 86.21%, and then, it declined. The variation in FF may be attributed to the alteration in the reverse saturation current [57]. The PCE showed an upward trend with the increasing $\phi_{BC}$, attaining a maximum value of 25.69% at 5.9 eV. But the PCE remained invariable after 5.9 eV, which was taken as the optimum $\phi_{BC}$. Therefore, Se, with a $\phi_{BC}$ of 5.9 eV, can be a probable substitution of Au to enhance the device performance. As the Schottky barrier diminishes at the higher $\phi_{BC}$ that causes a reduction in the series resistance, the PCE improves. However, the ohmic resistance increases at the interface HTL/back contact with the increasing $\phi_{BC}$, and consequently, the performance gets saturated [47].

### 3.12 Effect of the change in temperature

As the solar cells are usually set up in outdoor conditions, they face constant illumination from the sun. So, their temperature can become quite high compared to room temperature [58]. So, it is essential to figure out the impact of temperature on the PV parameters of PSCs.

Herein, the operating temperature was modulated from 300 K to 540 K. Fig. S10(a-d) of the Supplementary File shows the photovoltaic parameters trends against the operating temperature. The $V_{OC}$ attenuated from 1.0267 V to 0.7392 V because of the increment in reverse saturation current density ($J_0$) at the higher temperature, and the inverse relationship between the $V_{OC}$ and $J_0$. The correlation between them can be observed in equation (2):

$$V_{oc} = \frac{AK_B T}{q}\left[ln\left(1 + \frac{J_{SC}}{J_0}\right)\right] \dots\dots\dots\dots (2)$$

Here, $A$ stands for the ideality factor, and $K_B T/q$ represents the thermal voltage. Moreover, the defects enhance with the increasing temperature, which in turn, decrease the $V_{OC}$ [4].

For the PSC, there was hardly any change that could be noticed for $J_{SC}$. But at the higher temperature, the values of FF and PCE deteriorated drastically from 83.53% to 72.72% and 21.21% to 13.30%, respectively, which can be due to the reduction in shunt resistance [58]. In addition, the resistance in charge transfer rises with the temperature, thus deteriorating the charge recombination resistance and enhancing the recombination of charge carriers. Moreover, the other reasons for the deterioration of PCE with the augmentation of temperature can be the increment of ohmic resistance at the interfaces of different layers and the declination of the interfacial photo-carrier extraction in the PSC. To facilitate the PCE of PSC, additives can be introduced in HTL while performing fabrication [47]. The simulation result depicts that the PCE was highest at the room temperature of 300 K, which was taken as the optimum temperature for the PSC. This optimum temperature was comparable to the previously published works [4], [47].

**3.13 Effect of the change in HTL hole mobility**

According to the definition, the hole mobility ($\mu_h$) represents how a hole propagates under an electric field. The $\mu_h$ of an HTL is influenced by its $N_A$ and doping level. At a high acceptor doping level, the ionized impurity scattering limits the $\mu_h$. But at a low acceptor doping level, the lattice scattering limits the $\mu_h$ [2]. To scrutinize the impact of the $\mu_h$ of CuSbS$_2$ on the PV parameters, the $\mu_h$ was varied from $10^{-4}$ cm$^2$V$^{-1}$s$^{-1}$ to $10^2$ cm$^2$V$^{-1}$s$^{-1}$. Fig. S11(a) shows the PCE trends against the $\mu_h$ of CuSbS$_2$, whereas Fig. S11(b) depicts the J-V curves with different hole mobilities in the Supplementary File.

As the $\mu_h$ of CuSbS$_2$ was augmented, the J$_{SC}$ remained almost invariable, whereas the V$_{OC}$ augmented drastically from 0.6854 V to 1.0317 V [59]. The PCE significantly improved from 12.65% to 21.33% by enhancing the $\mu_h$ of CuSbS$_2$ because of the upgradation of hole conduction via the p-type layer. As the CuSbS$_2$ layer has elevated $N_A$ and $\mu_h$, its presence facilitates the PCE of the PSC [35]. Furthermore, a low $\mu_h$ is a reason behind the high series resistance, which in turn, causes deterioration in PCE [43]. The $\mu_h$ of the CuSbS$_2$ ($10^2$ cm$^2$V$^{-1}$s$^{-1}$) corresponding to the maximum PCE was taken as the optimum value.

The finally optimized J-V and QE curves have been shown in Fig. 12(a) and Fig. 12(b), respectively. After the final optimization, the device achieved a V$_{OC}$ of 1.1106 eV, J$_{SC}$ of 29.60 mA/cm$^2$, FF of 88.58%, and PCE of 29.13%. Therefore, the PCE of the device significantly improved by 7.92% in magnitude from the initial value of 21.21%. Table 6 depicts the comparison of the optimized device with the previous computational and experimental works on Cs$_2$TiBr$_6$ and MAPbI$_3$-based PSCs. Unlike MAPbI$_3$, Cs$_2$TiBr$_6$ has been infrequently studied as the absorber layer. It can be seen from Table 6 that our optimized device's performance was the best among all other existing PSCs. The absence of Pb made our device eco-friendly, and the use of all inorganic charge transport layers enforced its stability in the environment.

## 4. Conclusion

In this paper, we carried out a simulation study of an environmentally benign, non-toxic, and fully inorganic $Cs_2TiBr_6$-based PSC. Lead-free $Cs_2TiBr_6$ double perovskite was employed as the absorber layer. Moreover, for the very first time, LBSO was explored as the ETL for the simulation investigations of the PSC structure, replacing the commonly used $TiO_2$. To facilitate charge transportation, inorganic $CuSbS_2$ was used as the HTL instead of the traditionally used organic spiro-OMeTAD. The first-principle DFT calculations were performed to validate the 1.6 eV bandgap of the $Cs_2TiBr_6$ perovskite. In addition, we scrutinized the impact of various modulators on the PSC's PV parameters via SCAPS 1-D. The PSC showed the highest PCE at 1000 nm absorber thickness. However, we did not find any significant changes in PCE while modulating the thickness of LBSO. We finalized 400 nm thick $CuSbS_2$ as the optimal HTL, and the optimum absorber defect density was chosen as $10^{13}$ cm$^{-3}$. The optimized values of the doping density for the $Cs_2TiBr_6$, $CuSbS_2$, and LBSO were found to be $10^{15}$, $10^{20}$, and $10^{21}$ cm$^{-3}$, respectively. Besides, the highest PCE occurred at the 1.4 eV absorber bandgap. The effect of VBO between the HTL and the absorber was investigated by changing the $\varkappa_e$ of $CuSbS_2$, and the simulation results showed the maximum PCE at 4.1 eV $\varkappa_e$. Furthermore, different HTLs were investigated for the PSC by matching up the $V_{bi}$ with the $V_{OC}$. The $V_{bi}$ was found directly proportional to the $V_{OC}$, and the originally used $CuSbS_2$ showed the best PCE among all the HTLs. The optimum $\phi_{BC}$ was selected as 5.9 eV for the PSC, suggesting that Se can be a good replacement for Au as the back contact. Moreover, we identified that the PCE decreased with the increasing temperature, while it enhanced with the augmentation of the $\mu_h$ of HTL. Finally, the fully optimized device attained a PCE of 29.13%, which was a drastic improvement over the initial PCE of 21.21%. In addition, the validation of the SCAPS simulation was accomplished via wxAMPS. Therefore, the numerical simulation study of this proposed device will deliver a perspective for the development of a highly efficient, eco-friendly, and lead-free PSC.

## Acknowledgment

The authors would like to thank Dr. M. Burgelman at the University of Gent, Belgium, for issuing the SCAPS-1D software. They also would like to acknowledge Prof. A. Rockett and Dr. Yiming Liu from UIUC and Prof. Fonash of PSU for delivering the wxAMPS software.

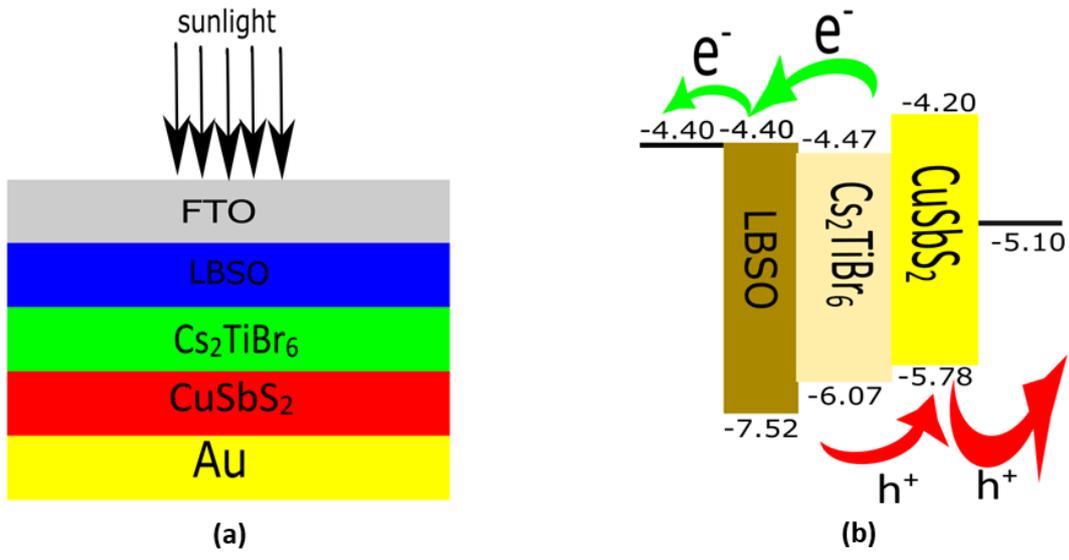

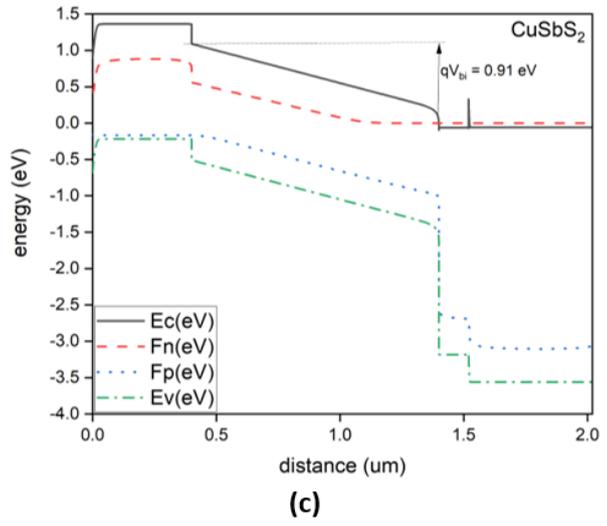

**Fig. 1. (a)** Schematic diagram, **(b)** Energy band alignment diagram, and **(c)** Energy band diagram of FTO/LBSO/$Cs_2TiBr_6$/$CuSbS_2$/Au.

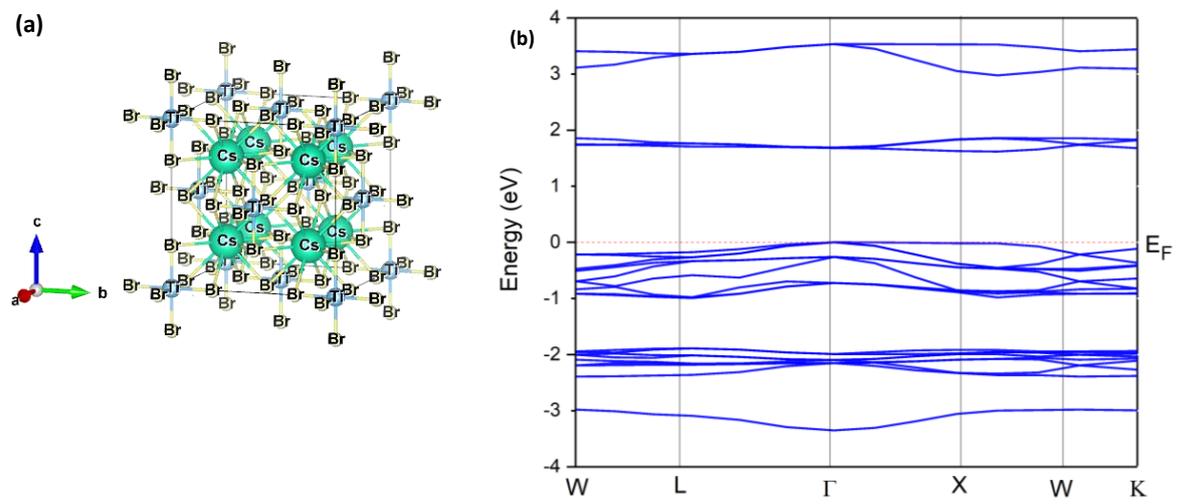

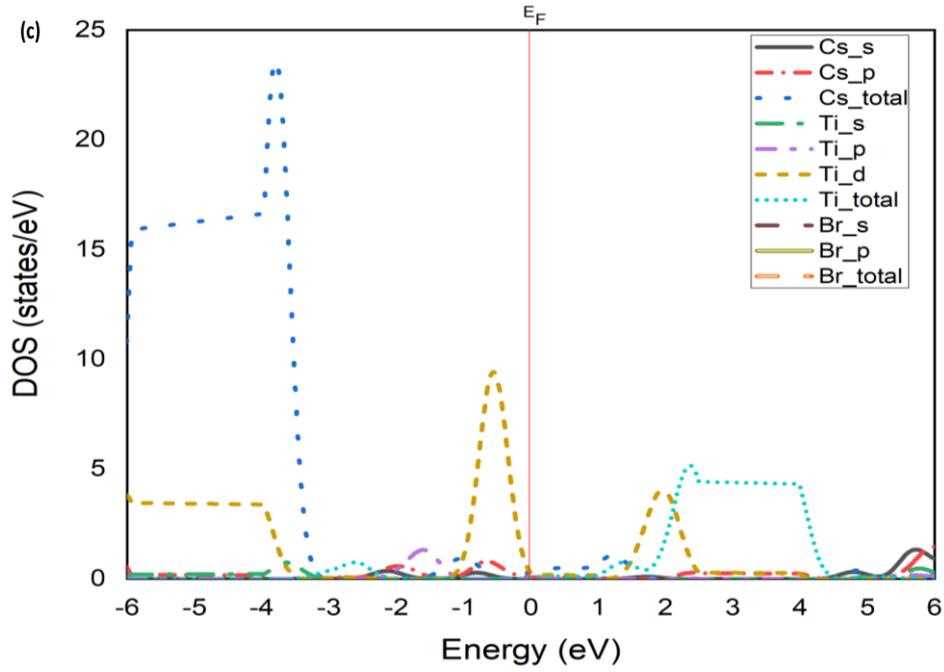

**Fig. 2. (a)** Unit cell of $Cs_2TiBr_6$ structure, **(b)** Band structure profile of $Cs_2TiBr_6$, and **(c)** DOS profile of $Cs_2TiBr_6$.

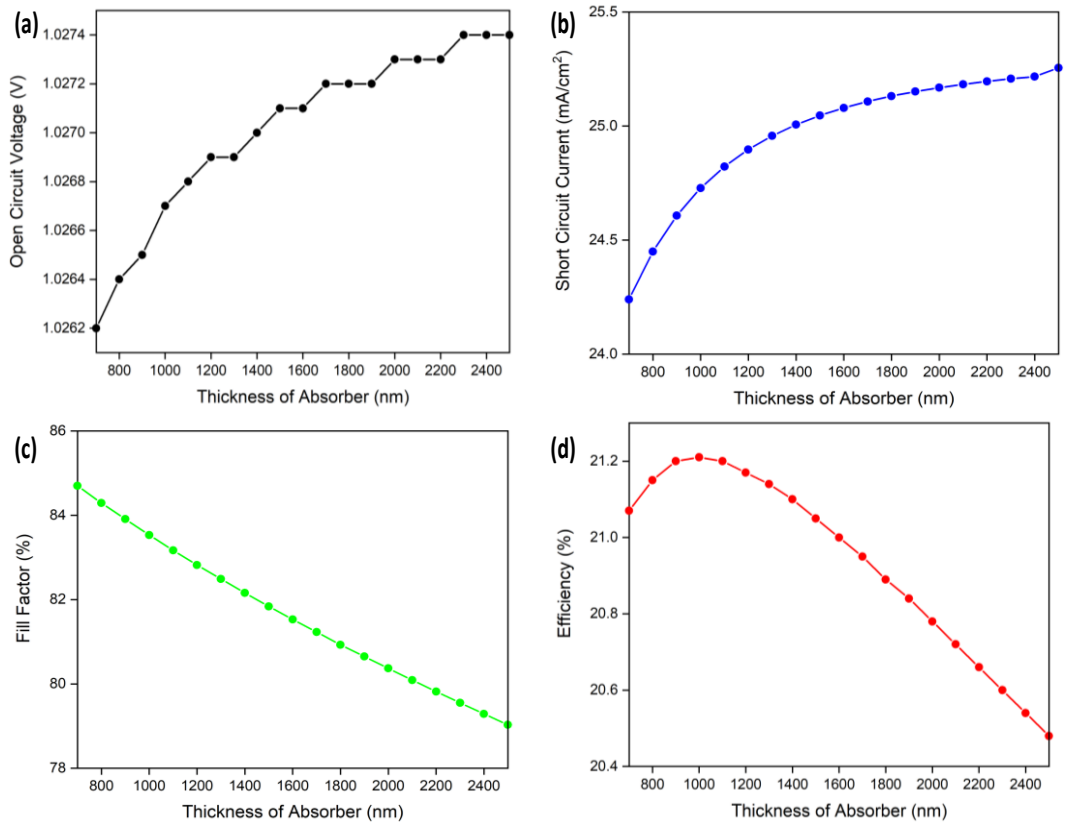

**Fig. 3.** Effect of the change in absorber thickness on **(a)** $V_{OC}$, **(b)** $J_{SC}$, **(c)** FF, and **(d)** PCE.

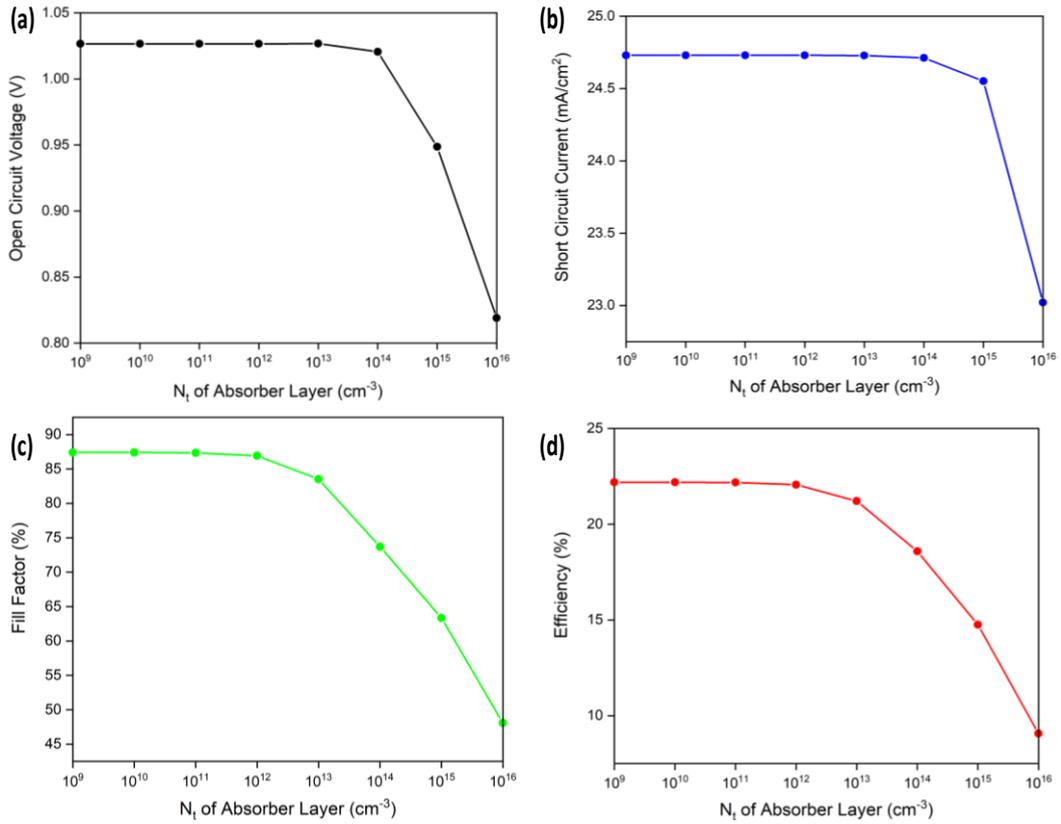

**Fig. 4.** Effect of the change in absorber layer defect density on **(a)** $V_{OC}$, **(b)** $J_{SC}$, **(c)** FF, and **(d)** PCE.

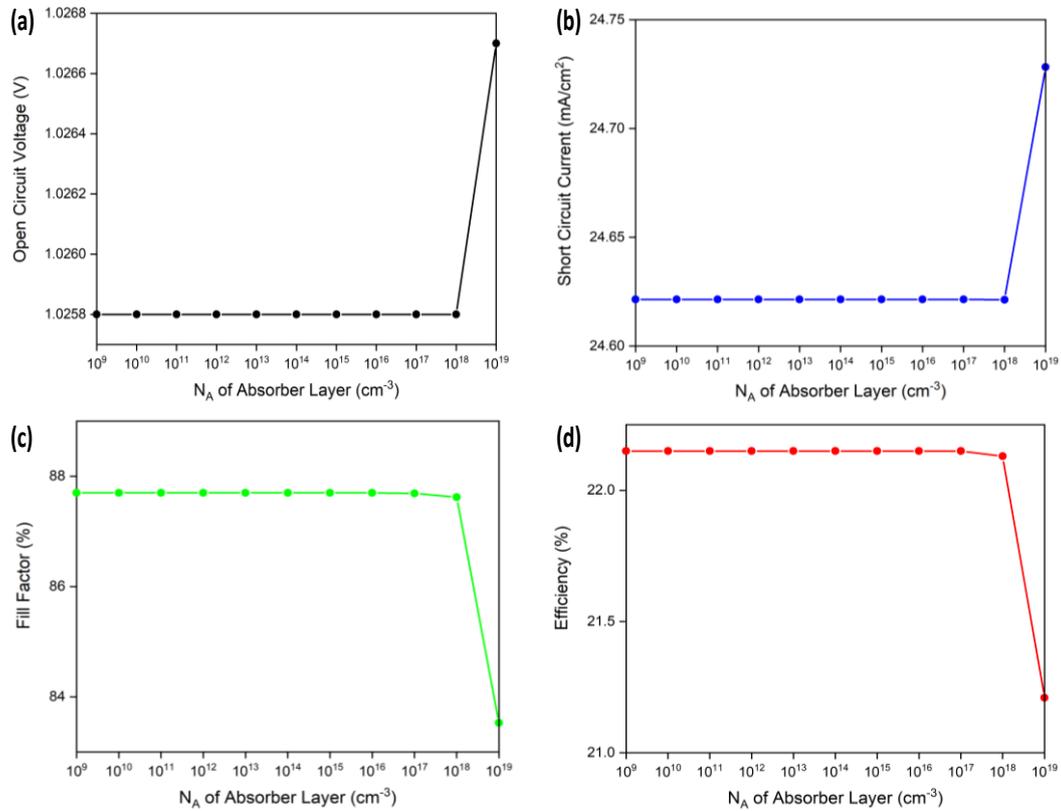

**Fig. 5.** Effect of the change in absorber doping density on **(a)** $V_{OC}$, **(b)** $J_{SC}$, **(c)** FF, and **(d)** PCE.

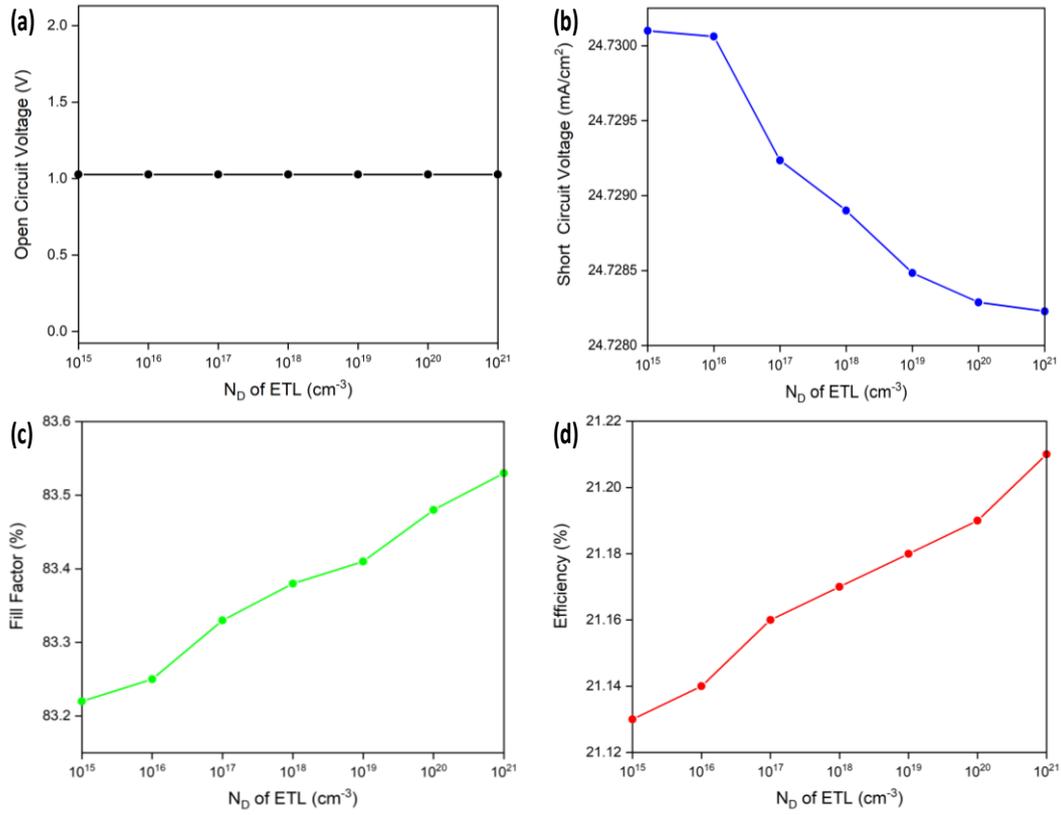

**Fig. 6.** Effect of the change in ETL doping density on **(a)** $V_{OC}$, **(b)** $J_{SC}$, **(c)** FF, and **(d)** PCE.

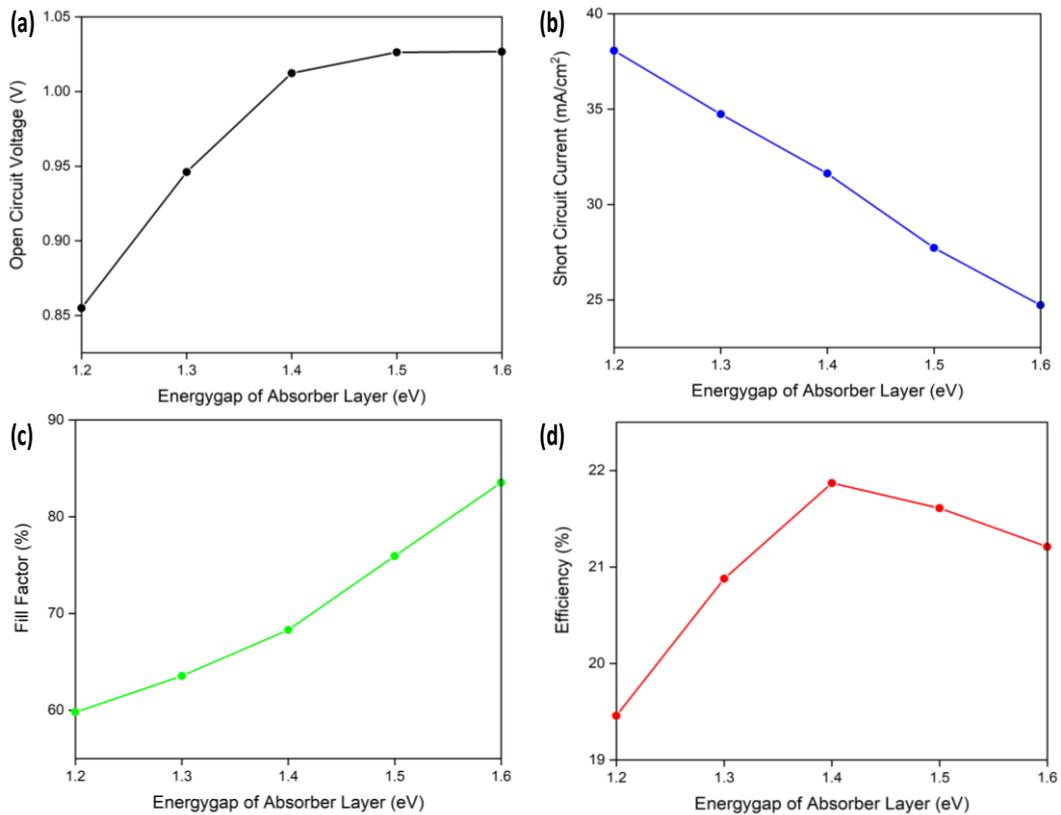

**Fig. 7.** Effect of the change in absorber layer energy gap on **(a)** $V_{OC}$, **(b)** $J_{SC}$, **(c)** FF, and **(d)** PCE.

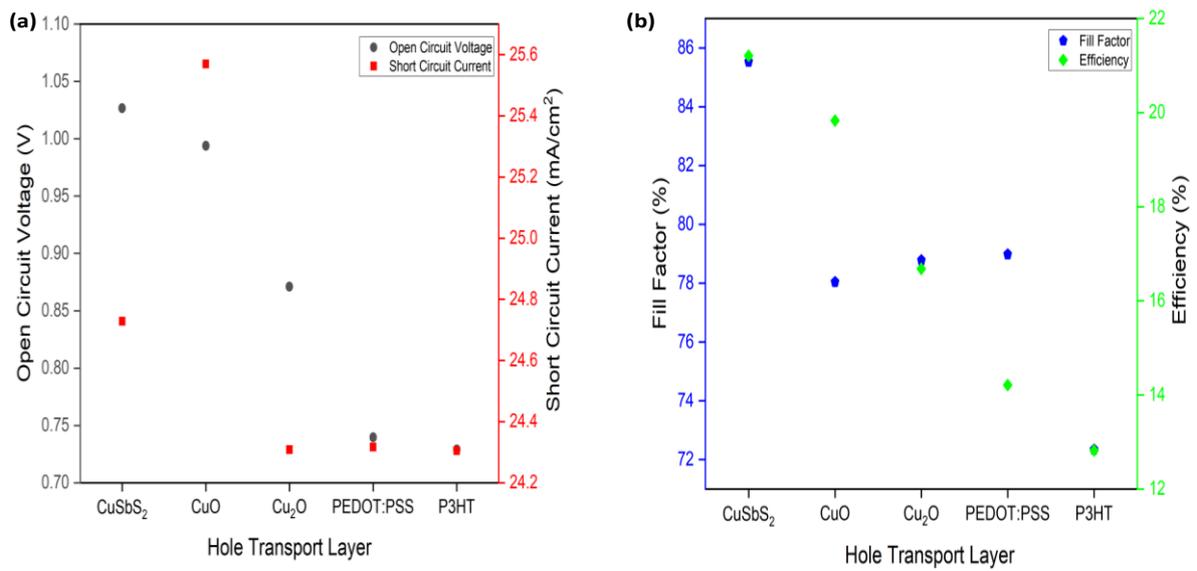

**Fig. 8.** Simulated **(a)** $V_{OC}$ and $J_{SC}$, **(b)** FF and PCE values for different HTLs of the PSC.

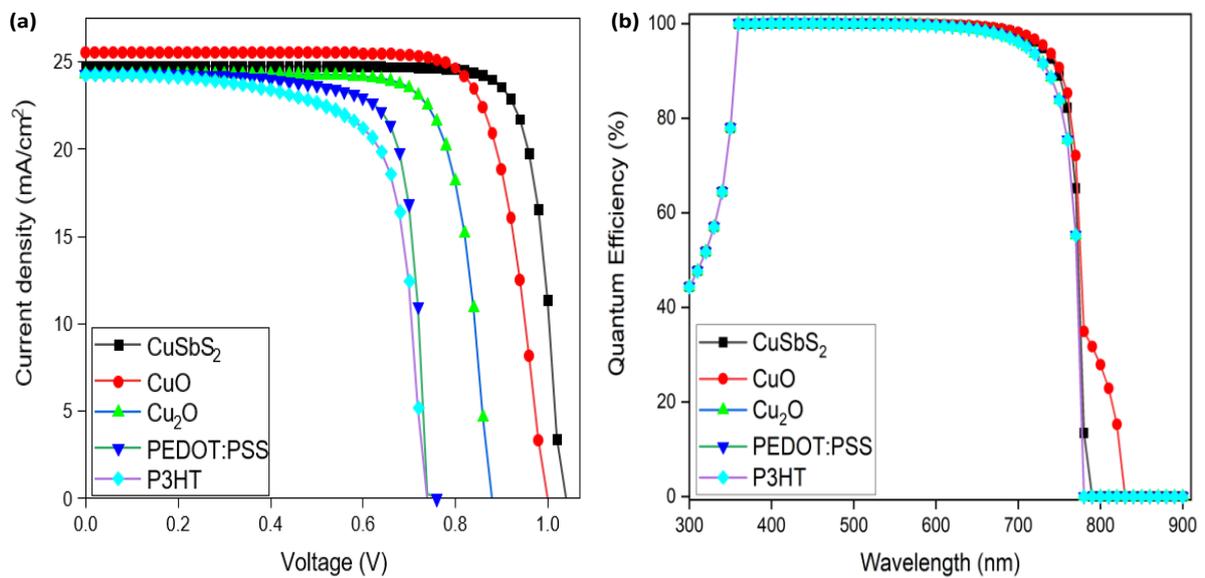

**Fig. 9.** Simulated **(a)** J-V and **(b)** QE curves for different HTLs of the PSC.

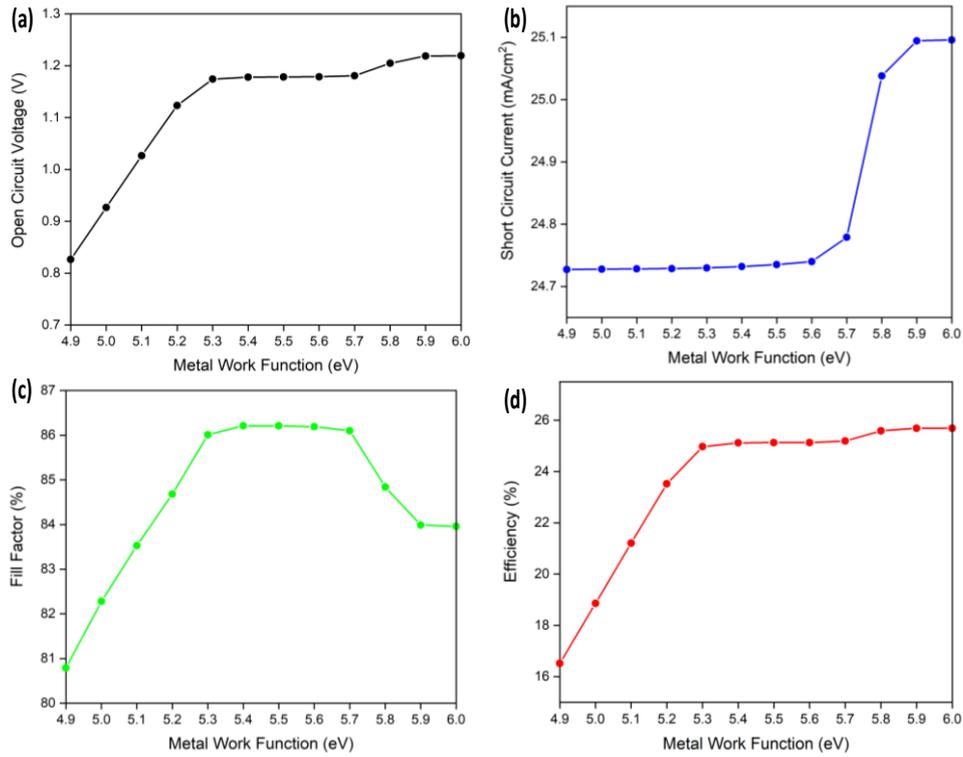

**Fig. 10.** Effect of the change in back contact work function on **(a)** $V_{OC}$, **(b)** $J_{SC}$, **(c)** FF, and **(d)** PCE.

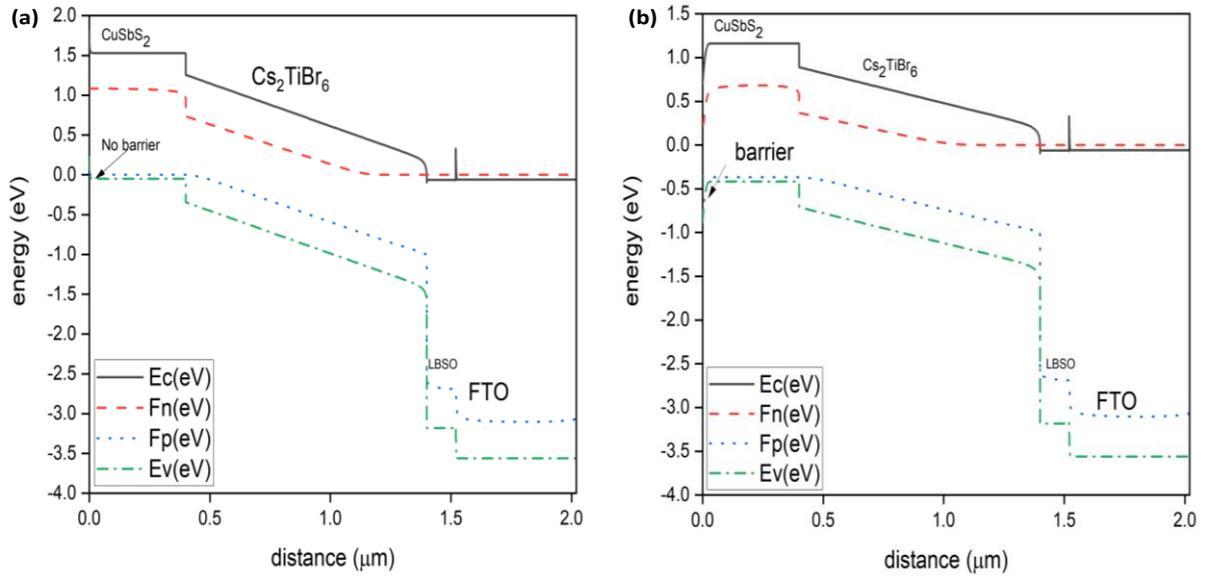

**Fig. 11.** Simulated energy band diagram for the metal work function of **(a)** 4.9 eV and **(b)** 6.0 eV.

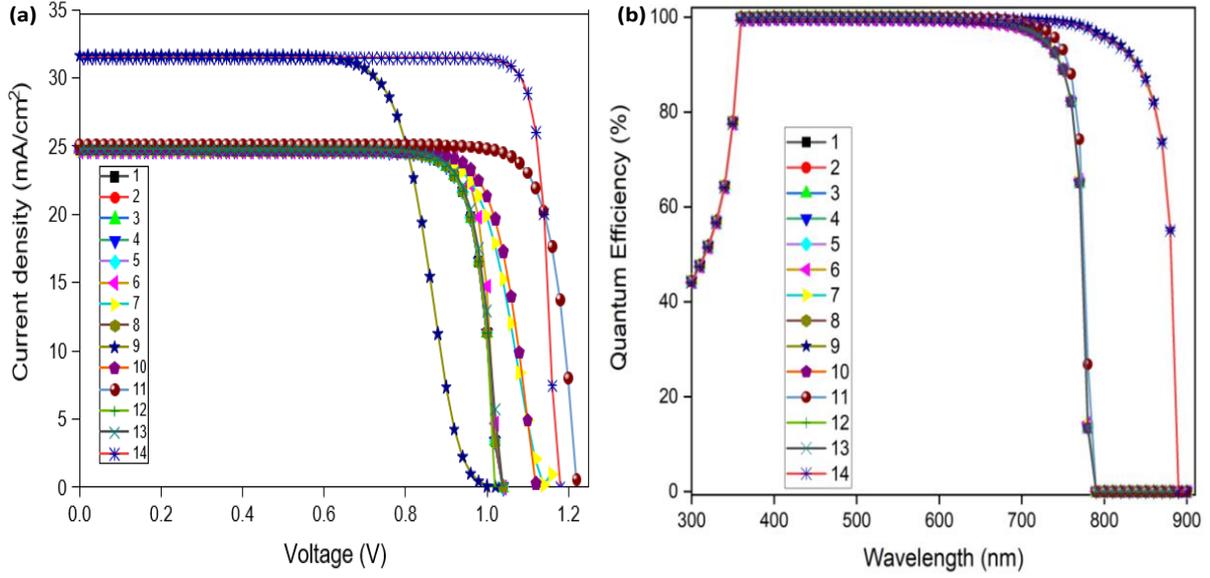

**Fig. 12.** Simulated optimized **(a)** J-V and **(b)** QE curves of the PSC (where the curve represents: **(1)** Initial parameters, **(2)** Optimized thickness of PAL, **(3)** Optimized thickness of ETL, **(4)** Optimized thickness of HTL, **(5)** Optimized $N_t$ of the absorber, **(6)** Optimized $N_A$ of the absorber, **(7)** Optimized $N_A$ of HTL, **(8)** Optimized $N_D$ of ETL, **(9)** Effect of $E_g$ of the absorber, **(10)** Effect of $\varkappa_e$ of HTL, **(11)** Effect of ɸ of back contact, **(12)** Effect of temperature, **(13)** Effect of $\mu_h$ of HTL, and **(14)** Final optimization).

**Table 1.** Basic parameters for charge transport layers and absorbers.

| Properties | Symbol | Unit | FTO | LBSO | $Cs_2TiBr_6$ | $CuSbS_2$ |
|---|---|---|---|---|---|---|
| thickness | W | nm | 500 | 120 | 1000 | 400 |
| bandgap | $E_g$ | eV | 3.5 | 3.12 | 1.6 | 1.58 |
| electron affinity | $\varkappa$ | eV | 4 | 4.4 | 4.47 | 4.2 |
| dielectric permittivity (relative) | $\epsilon$ | - | 9 | 22 | 10 | 14.6 |
| CB effective density of states | $N_c$ | $cm^{-3}$ | $2.2 \times 10^{18}$ | $1.8 \times 10^{20}$ | $6.0 \times 10^{19}$ | $2.0 \times 10^{18}$ |
| VB effective density of states | $N_v$ | $cm^{-3}$ | $1.8 \times 10^{19}$ | $1.8 \times 10^{20}$ | $2.14 \times 10^{19}$ | $1.0 \times 10^{19}$ |
| electron thermal velocity | $V_{thn}$ | cm/s | $1 \times 10^7$ | $1 \times 10^7$ | $1 \times 10^7$ | $1 \times 10^7$ |
| hole thermal velocity | $V_{thp}$ | cm/s | $1 \times 10^7$ | $1 \times 10^7$ | $1 \times 10^7$ | $1 \times 10^7$ |
| electron mobility | $\mu_n$ | $cm^2$/Vs | 20 | 0.69 | 4.4 | 49 |
| hole mobility | $\mu_p$ | $cm^2$/Vs | 10 | 0.69 | 2.5 | 49 |
| shallow uniform donor density | $N_D$ | $cm^{-3}$ | $2 \times 10^{19}$ | $2 \times 10^{21}$ | $1 \times 10^{19}$ | 0 |
| shallow uniform acceptor density | $N_A$ | $cm^{-3}$ | 0 | 0 | $1 \times 10^{19}$ | $1.38 \times 10^{18}$ |
| references | | | [28] | [23], [51], [60], [61] | [9], [62] | [15] |

**Table 2.** Defect parameters for charge transport layers and absorbers.

| Parameters | Symbol | Unit | FTO | LBSO | $Cs_2TiBr_6$ | $CuSbS_2$ |
|---|---|---|---|---|---|---|
| defect type | - | - | Neutral | Neutral | Neutral | Neutral |
| capture cross section electrons | $\sigma_n$ | $cm^2$ | $1 \times 10^{-15}$ | $1 \times 10^{-15}$ | $1 \times 10^{-15}$ | $1 \times 10^{-15}$ |
| capture cross section holes | $\sigma_p$ | $cm^2$ | $1 \times 10^{-15}$ | $1 \times 10^{-15}$ | $1 \times 10^{-15}$ | $1 \times 10^{-15}$ |
| energetic distribution | - | - | Single | Single | Single | Single |
| reference for defect energy level $E_t$ | - | - | Above $E_v$ | Above $E_v$ | Above $E_v$ | Above $E_v$ |
| energy level with respect of reference | - | eV | 0.6 | 0.6 | 0.7 | 0.1 |
| $N_t$ total | $N_t$ | $cm^{-3}$ | $1 \times 10^{14}$ | $1 \times 10^{14}$ | $1 \times 10^{13}$ | $1 \times 10^{14}$ |

**Table 3.** Defect parameters for interfacial contacts.

| Parameters | Symbol | Unit | $CuSbS_2/Cs_2TiBr_6$ | $Cs_2TiBr_6$/LBSO |
|---|---|---|---|---|
| defect type | - | - | Neutral | Neutral |
| capture cross section electrons | $\sigma_n$ | $cm^2$ | $1 \times 10^{-19}$ | $1 \times 10^{-19}$ |
| capture cross section holes | $\sigma_p$ | $cm^2$ | $1 \times 10^{-19}$ | $1 \times 10^{-19}$ |
| energetic distribution | - | - | gauß | gauß |
| reference for defect energy level $E_t$ | - | - | Above the highest EV | Above the highest EV |
| energy level w.r.t reference | - | eV | 0.6 | 0.6 |
| total density (integrated overall energies) | $N_t$ | $cm^{-2}$ | $3.90 \times 10^{10}$ | $3.90 \times 10^{10}$ |
| density at peak energy | - | $eVcm^{-2}$ | $2 \times 10^{10}$ | $2 \times 10^{10}$ |

**Table 4.** Basic parameters for different HTLs.

| Properties | Unit | CuO [63] | $Cu_2O$ [64] | PEDOT:PSS [10] | P3HT [55] |
|---|---|---|---|---|---|
| W | nm | 400 | 400 | 400 | 400 |
| $E_g$ | eV | 1.5 | 2.17 | 2.2 | 1.85 |
| $\varkappa$ | eV | 4.07 | 3.20 | 2.9 | 3.1 |
| $\epsilon$ | - | 18.1 | 7.11 | 3 | 3.4 |
| $N_c$ | $cm^{-3}$ | $2.0 \times 10^{18}$ | $2.0 \times 10^{18}$ | $2.0 \times 10^{18}$ | $2.0 \times 10^{18}$ |
| $N_v$ | $cm^{-3}$ | $1.0 \times 10^{19}$ | $1.0 \times 10^{19}$ | $1.0 \times 10^{19}$ | $1.0 \times 10^{19}$ |
| $V_{thn}$ | cm/s | $1.0 \times 10^7$ | $1.0 \times 10^7$ | $1.0 \times 10^7$ | $1.0 \times 10^7$ |
| $V_{thp}$ | cm/s | $1.0 \times 10^7$ | $1.0 \times 10^7$ | $1.0 \times 10^7$ | $1.0 \times 10^7$ |
| $\mu_n$ | $cm^2$/ V s | 100 | 80 | $2.0 \times 10^{-3}$ | $1.0 \times 10^{-4}$ |
| $\mu_p$ | $cm^2$/V s | 0.1 | 80 | $2.0 \times 10^{-3}$ | $1.0 \times 10^{-3}$ |
| $N_D$ | $cm^{-3}$ | 0 | 0 | 0 | 0 |
| $N_A$ | $cm^{-3}$ | $1.38 \times 10^{18}$ | $1.38 \times 10^{18}$ | $1.38 \times 10^{18}$ | $1.38 \times 10^{18}$ |

**Table 5.** Effect of $E_{C\_ETL}$-$E_{V\_HTL}$, $\phi_{BC}$ - $E_{C\_ETL}$, $qV_{bi}$, and photovoltaic parameters on the PSC.

| HTL | $\mu_p$ [$cm^2$/Vs] | $E_g$ [eV] | $E_{C\_ETL}$-$E_{V\_HTL}$ [eV] | $\phi_{BC}$ - $E_{C\_ETL}$ [eV] | $qV_{bi}$ [eV] | $V_{OC}$ [eV] | $J_{SC}$ [mA/$cm^2$] | FF [%] | PCE [%] |
|---|---|---|---|---|---|---|---|---|---|
| $CuSbS_2$ | 49 | 1.58 | 1.38 | 0.70 | 0.91 | 1.0267 | 24.728208 | 85.53 | 21.21 |
| CuO | 0.1 | 1.50 | 1.17 | 0.70 | 0.81 | 0.9939 | 25.569850 | 78.04 | 19.83 |
| $Cu_2O$ | 80 | 2.17 | 0.97 | 0.70 | 0.70 | 0.8711 | 24.308312 | 78.78 | 16.68 |
| PEDOT:PSS | $2.0 \times 10^{-3}$ | 2.20 | 0.70 | 0.70 | 0.48 | 0.7397 | 24.316510 | 78.98 | 14.21 |
| P3HT | $1.0 \times 10^{-3}$ | 1.85 | 0.55 | 0.70 | 0.29 | 0.7290 | 24.304812 | 72.34 | 12.82 |

**Table 6.** Performance comparison of existing $Cs_2TiBr_6$ and $MAPbI_3$-based PSCs.

| Structure | PCE [%] | $V_{OC}$ [V] | $J_{SC}$ (mA/cm²) | FF [%] | References |
|---|---|---|---|---|---|
| FTO/TiO$_2$/Cs$_2$TiBr$_6$/NiO/Au | 8.51 | 1.12 | 10.25 | 73.59 | [65] |
| FTO/SnO$_2$/ Cs$_2$TiBr$_6$/MoO$_3$/Au | 11.49 | 1.53 | 8.66 | 86.45 | [66] |
| FTO/V$_2$O$_5$/ Cs$_2$TiBr$_6$/CdTe/Au | 14.55 | 0.92 | 18.176 | 86.58 | [67] |
| FTO/ZnO/ Cs$_2$TiBr$_6$/Cu$_2$O/Au | 18.15 | 1.53 | 13.60 | 87.23 | [45] |
| FTO/TiO2/MAPbI$_3$/PTAA/Au | 19.6 | 1.07 | 23.3 | 78.6 | [23] |
| FTO/LBSO/MAPbI$_3$/PTAA/Au | 21.3 | 1.12 | 23.4 | 81.3 | [23] |
| FTO/LBSO/Cs$_2$TiBr$_6$/CuSbS$_2$/Se | 29.13 | 1.11 | 29.60 | 88.58 | This Work |

# Supplementary File

## Computational equations used in SCAPS:

The Poisson's equation (equation 1), the electron continuity equation (equation 2), and the hole continuity equation (equation 3) are given below [1]:

$$\frac{d}{dx}\left(-\varepsilon(x)\frac{d\psi}{dx}\right) = q[p(x) - n(x) + N_d^+(x) - N_a^-(x)] \ldots \ldots \ldots (1)$$

$$\frac{\partial j_n}{\partial x} = q\left(R_n - G + \frac{\partial n}{\partial t}\right) \ldots \ldots \ldots (2)$$

$$\frac{\partial j_p}{\partial x} = -q\left(R_p - G + \frac{\partial p}{\partial t}\right) \ldots \ldots \ldots (3)$$

Where, $\varepsilon$ is the permittivity, $q$ is the electron charge, $\psi$ is the electrostatic potential, $n$ is the total electron density, $p$ is the total hole density, $N_d^+$ is the ionized donor-like doping concentration, $N_a^-$ is the ionized acceptor-like doping concentration, $j_n$ and $j_p$ are the electron and hole current densities respectively, $R_n$ and $R_p$ are the net recombination rates for electron and hole per unit volume respectively, and $G$ is the generation rate per unit volume.

## Computational equations utilized in wxAMPS:

In 1-D space, Poisson's equation is given by [2]:

$$\frac{d}{dx}\left(-\varepsilon(x)\frac{d\psi\prime}{dx}\right) = q.[p(x) - n(x) + N_D^+(x) - N_A^-(x) + pt(x) - nt(x)] \ldots \ldots \ldots (4)$$

Where, the electrostatic potential $\psi'$ and the free electron n, free hole p, trapped electron nt, and trapped hole pt as well as the ionized donor-like doping $N_D^+$ and ionized acceptor-like doping $N_A^-$ concentrations are all functions of the position coordinate x. The continuity equation for the free electrons in the delocalized states of the conduction band has the form [3]:

$$\frac{1}{q}\left(\frac{dJn}{dx}\right) = -G_{op}(x) + Rx \quad \ldots \ldots \ldots (5)$$

Again, the continuity equation for the free holes in the delocalized states of the valence band has the form:

$$\frac{1}{q}\left(\frac{dJp}{dx}\right) = G_{op}(x) - Rx \quad \ldots \ldots \ldots (6)$$

Where, Jn and Jp are, respectively, the electron and hole current densities. The term R(x) is the net recombination rate resulting from band-to-band (direct) recombination and SRH (indirect)

recombination traffic through gap states. The net direct recombination rate is [4]:

$$R_D(x) = \beta(np - ni^2) \quad \ldots \ldots \ldots (7)$$

Where, $\beta$ is a proportionality constant, which depends on the material's energy band structure under analysis, and n and p are the band carrier concentrations present when devices are subjected to a voltage bias, light bias, or both. The continuity equations include the term $G_{op}(x)$, which is the optical generation rate as a function of x due to externally imposed illumination.

**Absorption data for the perovskite solar cell:**

Absorption data for each layer was achieved from the new Eg-sqrt model (SCAPS version 3.3.07), which is the updated model of the traditional SCAPS model (traditional sqrt (hυ-$E_g$) law model) and can be found from the "Tauc laws". The updated Eg-sqrt model follows equation 8 [5].

$$\alpha(h\upsilon) = (\alpha_0 + \beta_0 \frac{E_g}{h\upsilon})\sqrt{\frac{h\upsilon}{E_g} - 1} \ldots \ldots \ldots (8)$$

Where, $\alpha$ is the optical absorption constant, $h\upsilon$ is the photon energy, and $E_g$ is the bandgap. The model constants $\alpha_0$ and $\beta_0$ have the dimension of absorption constant (e.g., 1/cm) and are related to the traditional model constants A and B by the relations:

$$\alpha_0 = A\sqrt{E_g} \text{ and } \beta_0 = \frac{B}{\sqrt{E_g}}$$

**SRH recombination due to the defects in the perovskite layer:**

The charge carriers in the PSC are recombined by Shockley-Read-Hall (SRH) recombination process and the net recombination rate ($R^{SRH}$) for SRH recombination is given by the following equation [1]:

$$R^{SRH} = \frac{\upsilon \sigma_n \sigma_p N_T [np - n_i^2]}{\sigma_p [p + p_1] + \sigma_n [n + n_1]} \ldots \ldots \ldots (9)$$

Where, $\sigma_n$ and $\sigma_p$ are the capture cross-sections for electrons and holes, $\upsilon$ is the electron thermal velocity, $N_T$ is the atomistic defect concentration, $n_i$ is the intrinsic carrier density, $n$ and $p$ are the concentrations of electron and hole at equilibrium, and $n_1$ and $p_1$ are the concentrations of electrons and holes in trap defect and valence band, respectively.

According to equation 9, the $R^{SRH}$ is directly proportional to the defect density in the perovskite absorber layer. Again, the $R^{SRH}$ has an impact on the carrier diffusion length. The diffusion length increases with decreasing the perovskite absorber layer's defect density, which improves the solar cell performance. The relation between the diffusion length, carrier mobility, and lifetime at a temperature T is expressed in equation 10 [4].

$$L_D = \sqrt{\frac{\mu_{(e,h)} R^{SRH}_T}{q} \times \tau_{lifetime}} \ldots \ldots \ldots (10)$$

Where, $L_D$, $\mu_{(e,h)}$, and $\tau_{lifetime}$ are the diffusion length, the electron and hole mobility, and the minority-carrier lifetime, respectively. Moreover, $\tau_{lifetime}$ depends upon the defect density and capture cross-section area for electrons and holes. The relation between $\tau_{lifetime}$ and bulk defect density is expressed in equation 11.

$$\tau_{lifetime} = \frac{1}{N_T \delta v_{th}} \ldots \ldots \ldots (11)$$

Here, $\delta$, $v_{th}$, and $N_T$ represent the capture cross-section area for electrons and holes, the thermal velocity of carriers, and defect concentration, respectively.

**Validation of the model using wxAMPS:**

The simulations were carried out using wxAMPS (version 2.0) for validating the results attained from the SCAPS software. The simulations were run at 1000 nm absorber thickness in both software to find the variations in $V_{OC}$, $J_{SC}$, FF, and PCE of the PSC. All the simulations were performed at 300 K working temperature and AM1.5G solar spectrum. Table S1 shows the comparison in the photovoltaic performance between the two simulation tools. Furthermore, Fig. S1(a) and Fig. S1(b) exhibit the comparison in the J-V and QE curves, respectively, between the two simulation tools.

**Table S1.** Comparison between the SCAPS and wxAMPS results at 1000 nm absorber thickness of the PSC.

| Serial | Software | $V_{OC}$ V | $J_{SC}$ mA/cm² | FF % | PCE % |
|---|---|---|---|---|---|
| 1 | SCAPS | 1.0267 | 24.7282 | 83.53 | 21.21 |
| 2 | wxAMPS | 1.0165 | 23.8396 | 88.88 | 21.53 |

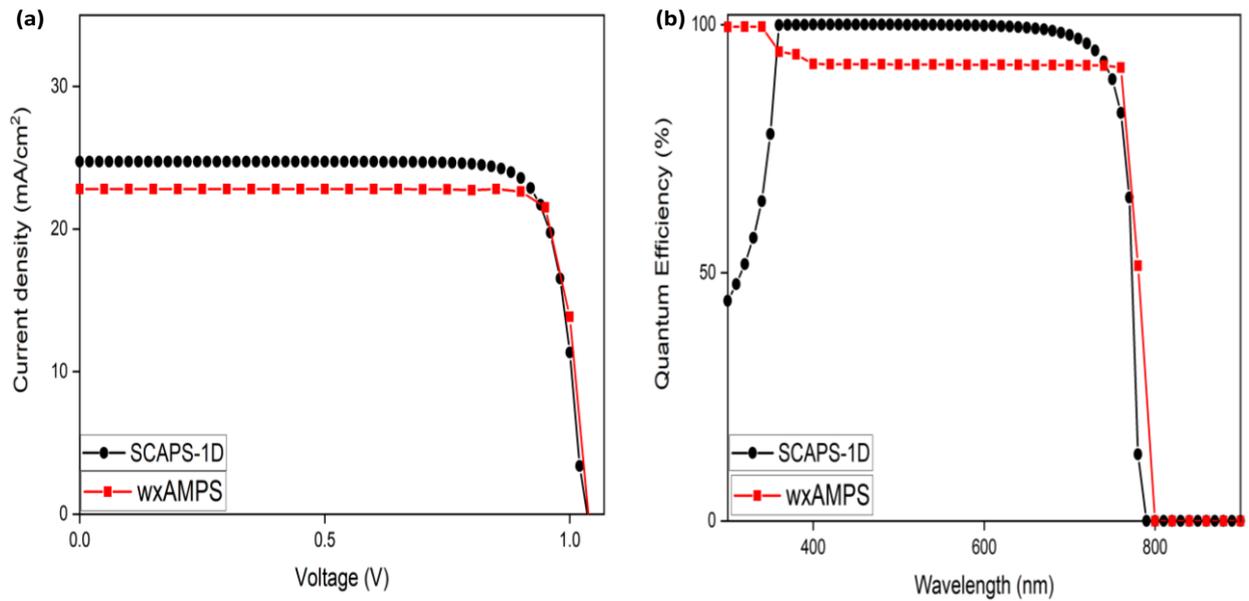

**Fig. S1.** Comparison in the **(a)** J-V and **(b)** QE curves of the PSC between the wxAMPS and SCAPS software at 1000 nm absorber thickness.

**SCAPS simulation working procedure:**

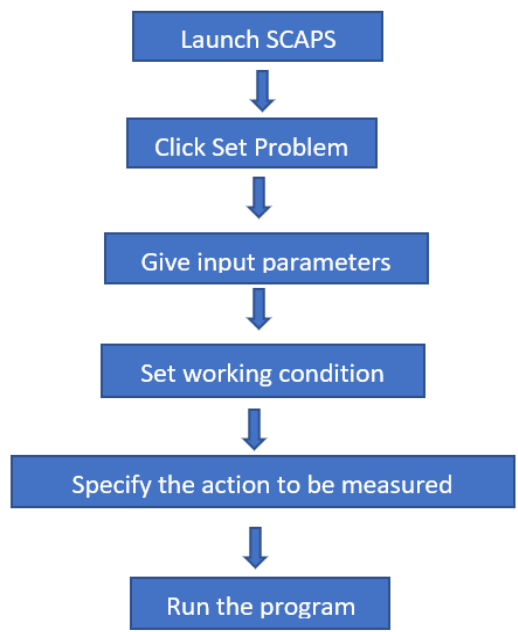

**Fig. S2.** SCAPS simulation working process [6].

**Results and Discussion:**

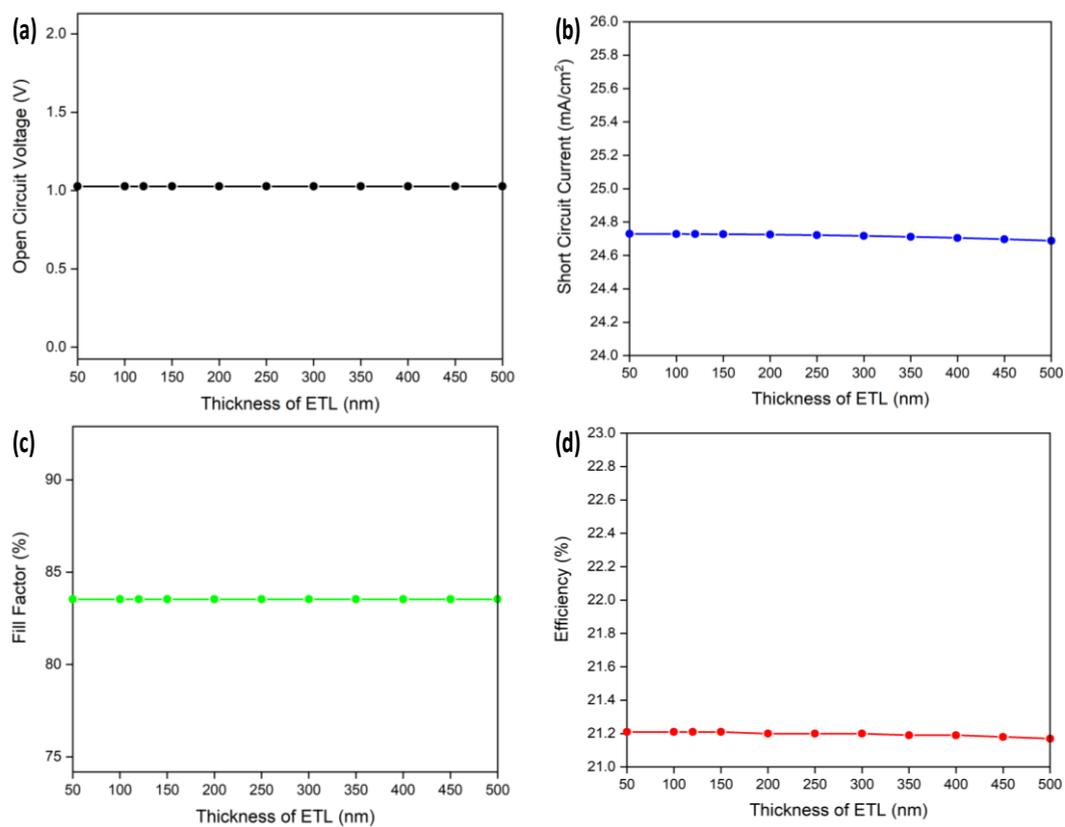

**Fig. S3.** Effect of the change in ETL thickness on **(a)** $V_{OC}$, **(b)** $J_{SC}$, **(c)** FF, and **(d)** PCE.

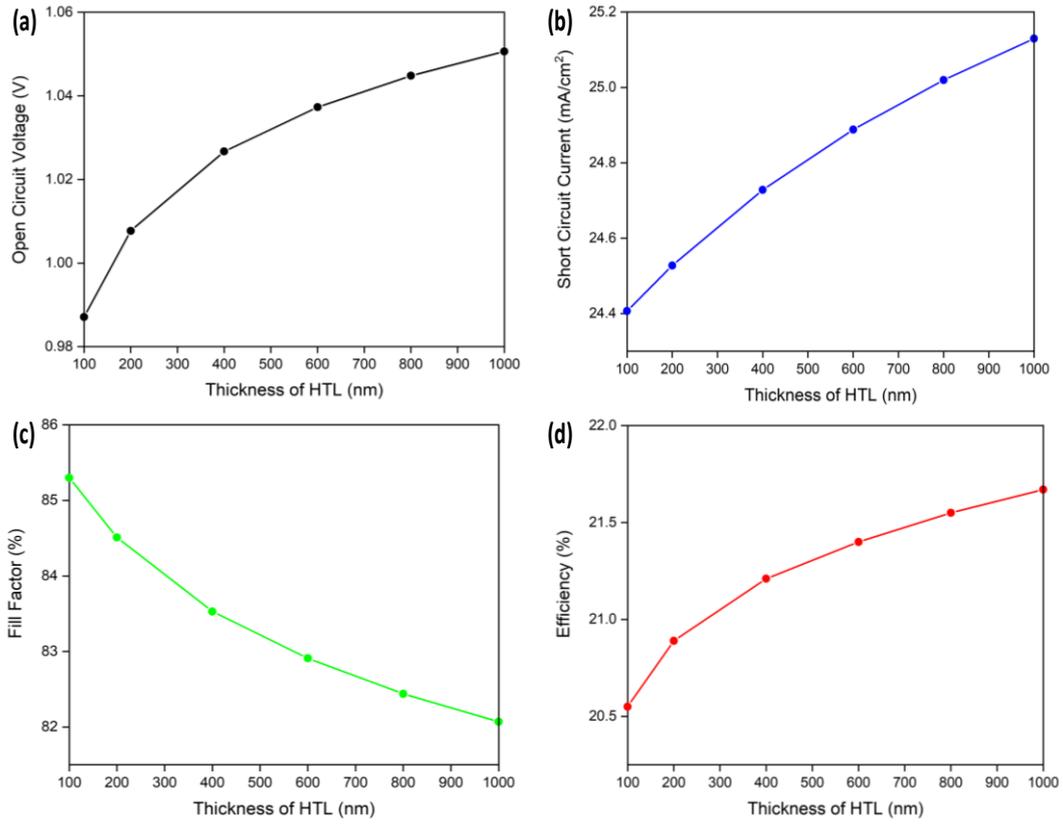

**Fig. S4.** Effect of the change in HTL thickness on **(a)** $V_{OC}$, **(b)** $J_{SC}$, **(c)** FF, and **(d)** PCE.

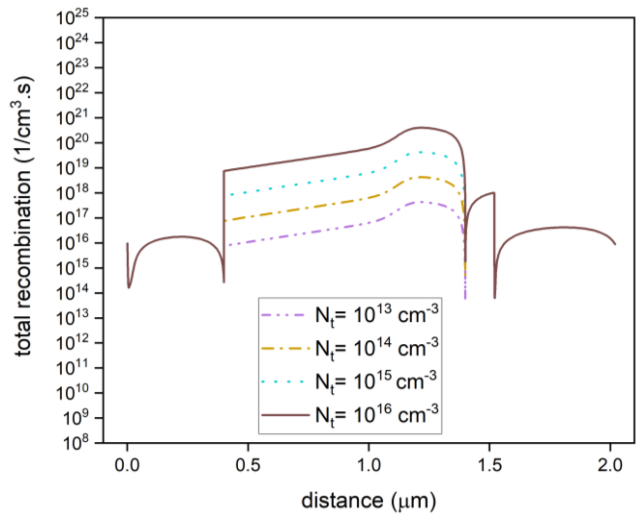

**Fig. S5.** Total recombination profile of absorber layer at different absorber defect densities.

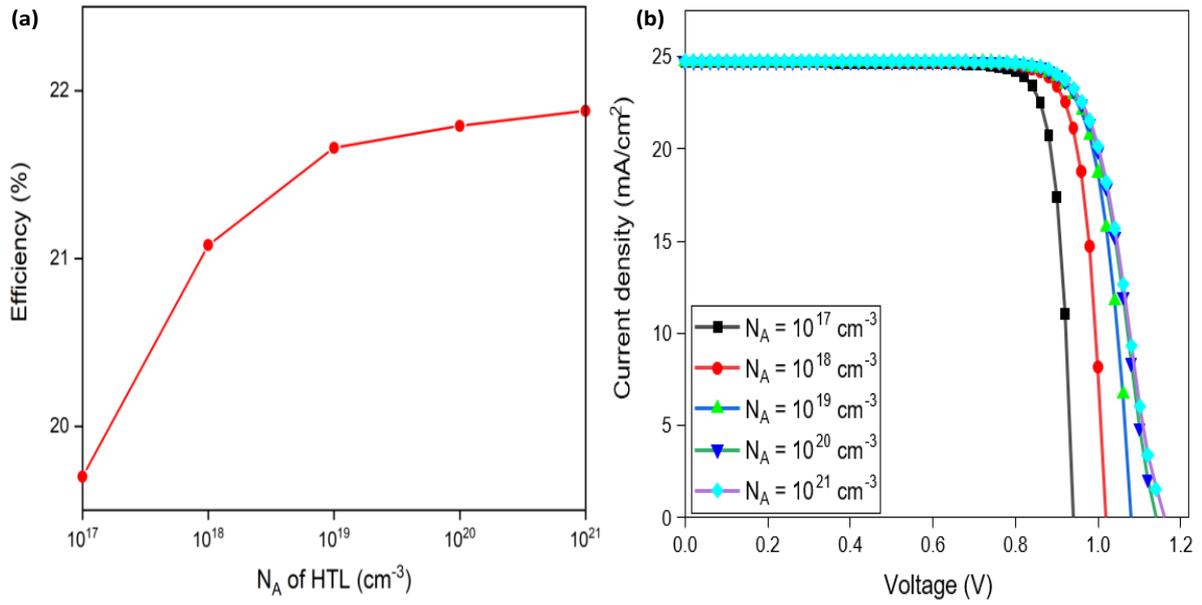

**Fig. S6.** Effect of the change in HTL doping density on **(a)** PCE and **(b)** J-V curves.

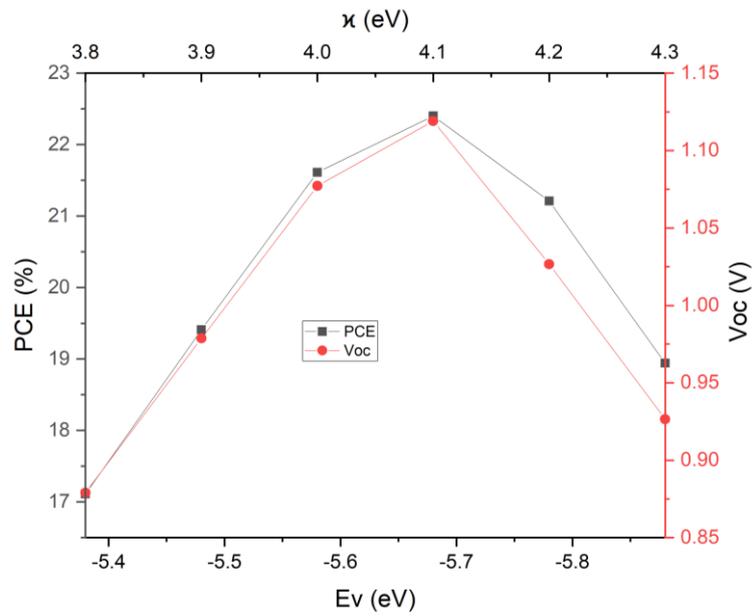

**Fig. S7.** Influence of the electron affinity of CuSbS$_2$ on the performance of the PSC.

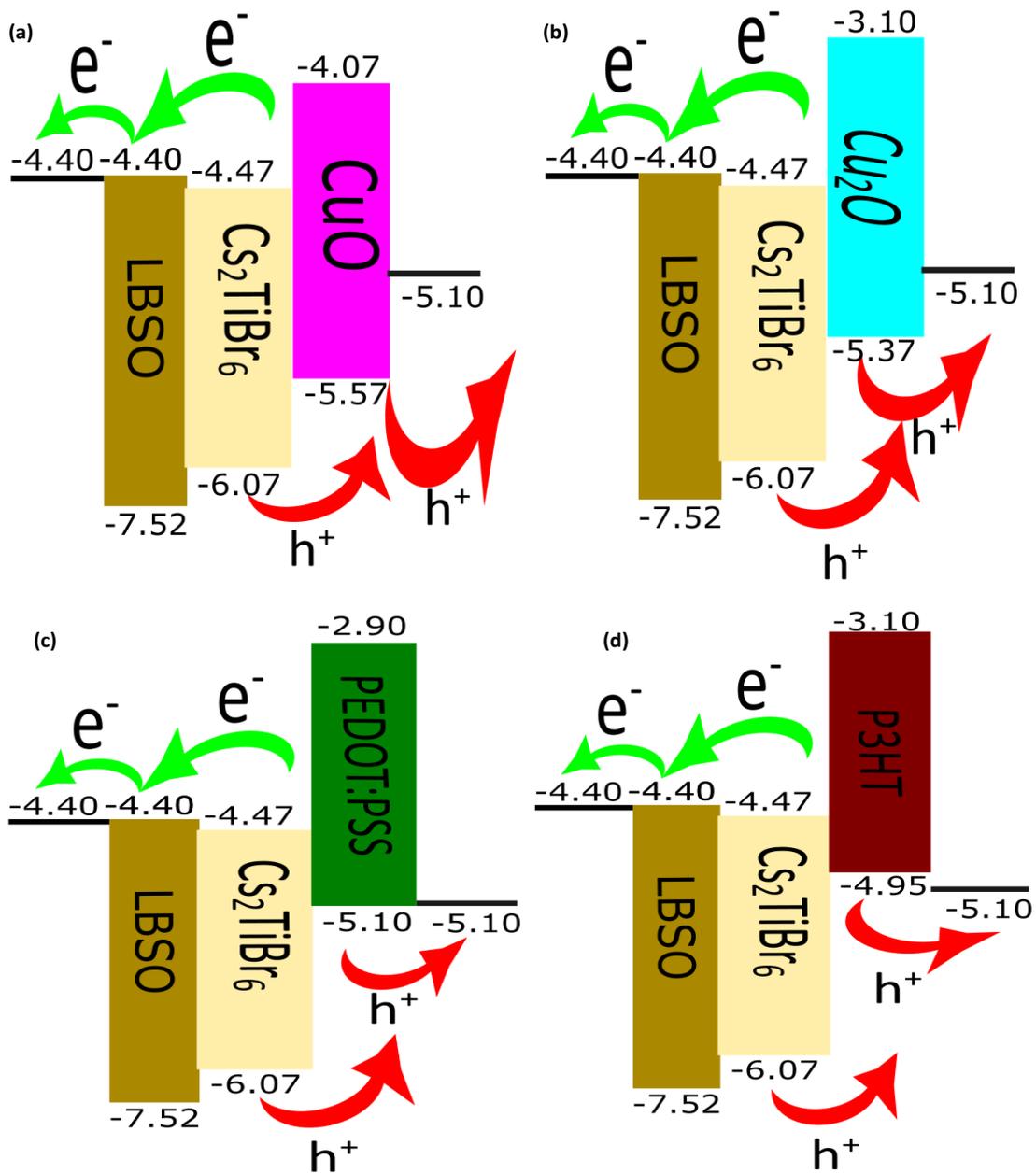

**Fig. S8.** Energy band alignment diagram for various HTLs of the PSC; **(a)** CuO, **(b)** Cu$_2$O, **(c)** PEDOT:PSS, and **(d)** P3HT.

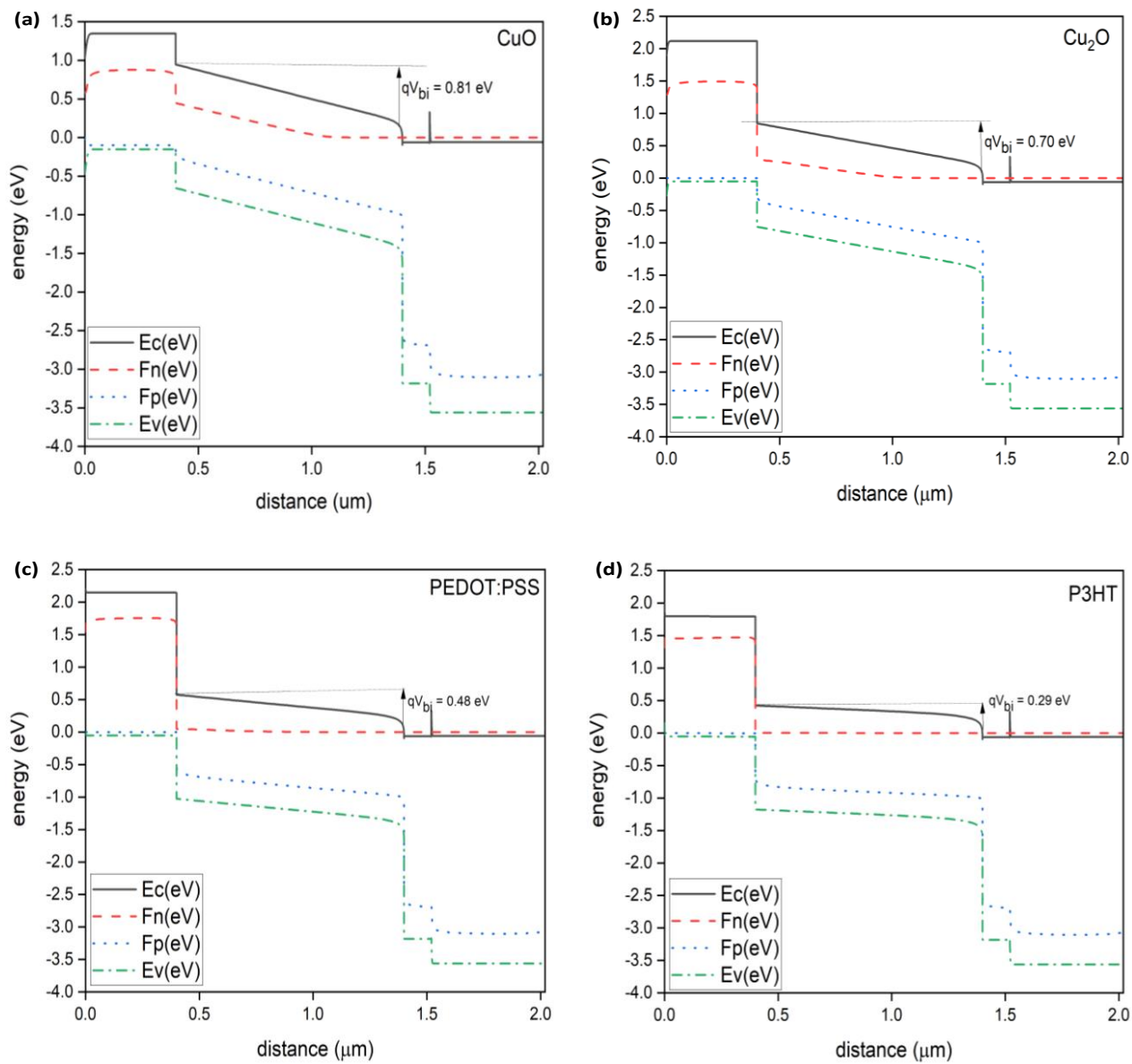

**Fig. S9.** Energy band diagram for various HTLs of the PSC; **(a)** CuO, **(b)** $Cu_2O$, **(c)** PEDOT:PSS, and **(d)** P3HT.

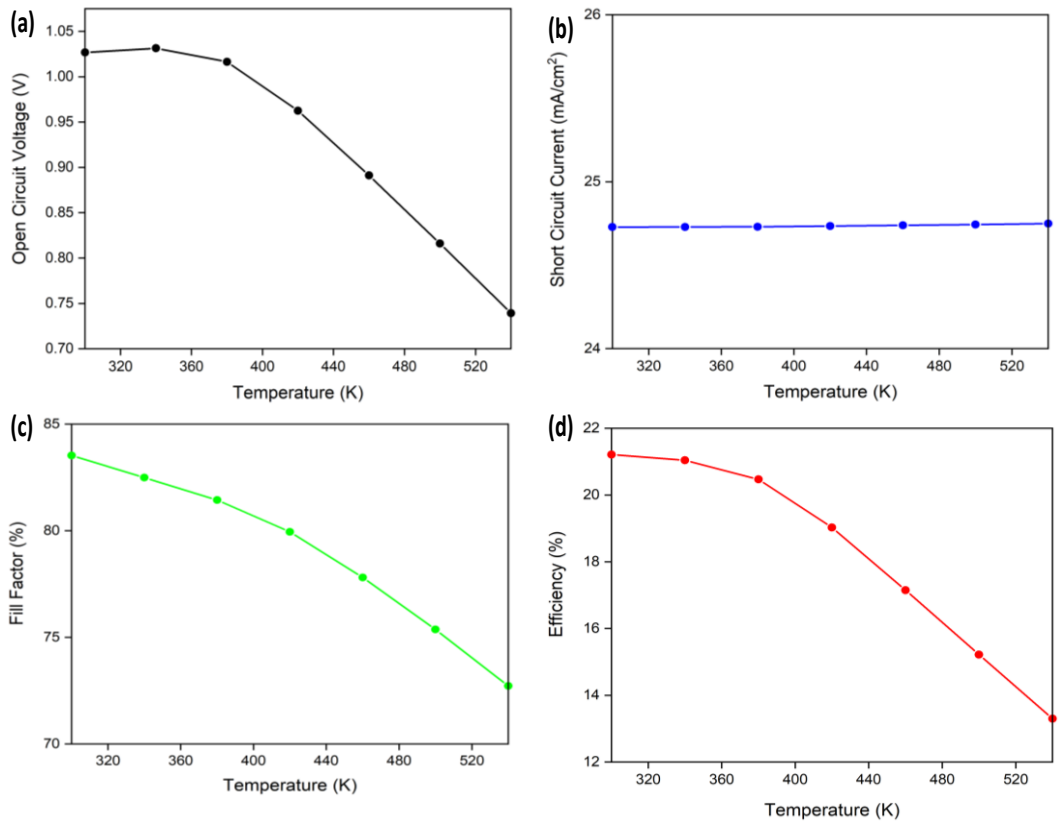

**Fig. S10.** Effect of the change in temperature on **(a)** $V_{OC}$, **(b)** $J_{SC}$, **(c)** FF, and **(d)** PCE.

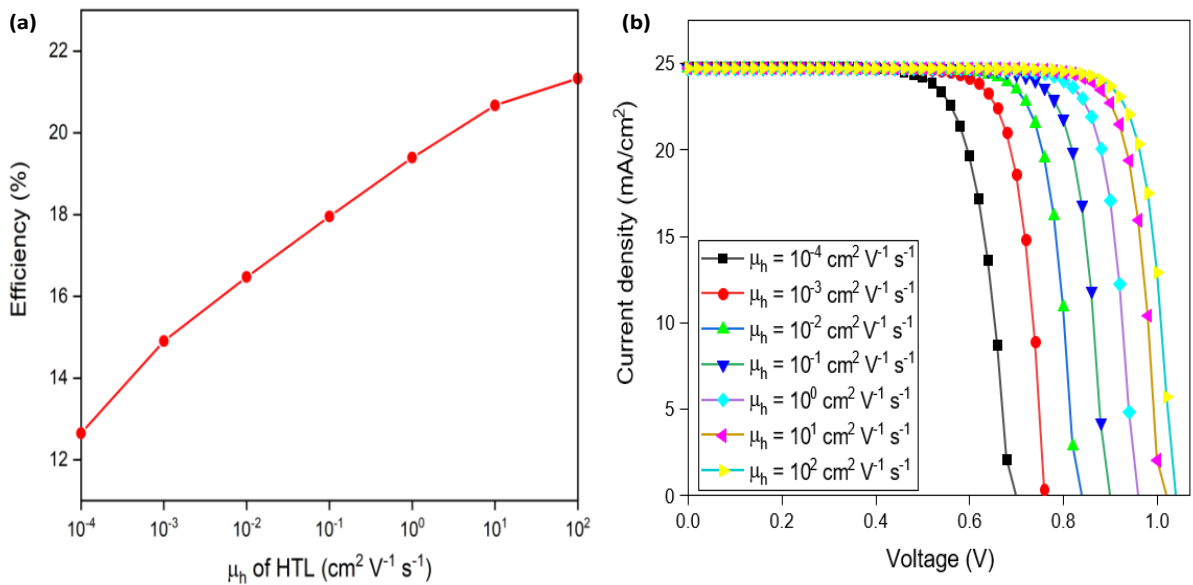

**Fig. S11.** Effect of the change in hole mobility of HTL on the **(a)** PCE and **(b)** J-V curves.